\documentclass[acmsmall]{acmart}
\usepackage{stmaryrd}
\usepackage[utf8]{inputenc}
\usepackage{alltt}
\usepackage{array}
\usepackage{mathpartir}
\usepackage{listings,lstcoq}
\usepackage{afterpage}
\newcommand{\evalf}[2]{\mathtt{eval}\ #1\ #2}
\newcommand{\reducef}[3]{\mathtt{reduce}\ #1\ #2\ #3}
\newcommand{\applyf}[2]{\mathtt{apply}\ #1\ #2}
\newcommand{\convfname}{\mathtt{conv}^?}
\newcommand{\convvfname}{\mathtt{conv}}
\newcommand{\cstkfname}{\mathtt{conv}^*}
\newcommand{\convf}[3]{\convfname\ #1\ #2\ #3}
\newcommand{\convvf}[3]{\convvfname\ #1\ #2\ #3}
\newcommand{\cstkf}[3]{\cstkfname\ #1\ #2\ #3}
\newcommand{\readbackf}[1]{\mathtt{reify}\ #1}
\newcommand{\normalizef}[1]{\mathtt{nf}\ #1}

\newcommand{\shallowclos}[3]{\langle #1, #2, #3 \rangle}
\newcommand{\neutralv}[2]{[#1\ #2]}
\newcommand{\neutralvar}[1]{[#1]}
\newcommand{\deepclos}[5]{\langle #1, #2, #3, #4, #5 \rangle}
\newcommand{\constv}[3]{[#1\ #2]@#3}
\newcommand{\constvinit}[2]{[#1]@#2}
\newcommand{\frozenconstv}[2]{[#1\ #2]}

\newcommand{\emptyenv}{\{\,\}}
\newcommand{\getenv}[2]{#1(#2)}
\newcommand{\extenv}[3]{#1 + #2 \mapsto #3}

\newcommand{\sendp}[2]{#1 \mathbin{!} #2}
\newcommand{\recvp}[1]{#1?}
\newcommand{\parp}[2]{#1 \parallel #2}
\newcommand{\parpp}[3]{#1 \parallel #2 \parallel #3}

\newcommand{\freshp}[2]{\nu #1 . \, #2}

\newcommand{\choice}{\mathrel{\oplus}}
\newcommand{\biasedor}{\mathrel{\overrightarrow{\oplus}}}

\newcommand{\trueb}{\mathsf{T}}
\newcommand{\falseb}{\mathsf{F}}

\newcommand{\red}{\rightarrow}
\newcommand{\redplus}{\mathrel{\red^+}}

\newcommand{\redbd}{\mathrel{\red}_{\beta\delta}}

\catcode`"=\active
\def"#1"{\hbox{\texttt{#1}}}         %
\def\lam#1.{\lambda #1.\,}            %
\newcommand{\convertible}{\approx}   %
\newcommand{\Dom}{\mathrm{Dom}}

\newcommand{\nil}{\epsilon}
\newcommand{\cons}{\cdot}
\newcommand{\append}{\cdot}
\newcommand{\listlength}[1]{|#1|}

\newcommand{\becomes}{:=}  %

\newcommand{\secref}[1]{\S\ref{#1}}

\newcommand{\schedstate}[1]{\big[#1\big]}
\newcommand{\machstate}[2]{#1~\schedstate{#2}}
\newcommand{\schedstatetwolines}[2]{\big[#1, \\ ~~#2\big]}
\newcommand{\machstatetwolines}[2]{\begin{array}[t]{@{}l@{}}#1 \\ \schedstate{#2} \end{array}}
\newcommand{\finish}{\mathbf{finish}}
\newcommand{\wakeup}{\mathbf{need}}
\newcommand{\unneed}{\mathbf{unneed}}
\newcommand{\upd}{\mapsto}
\newcommand{\updatew}[3]{#1 [ #2 \upd #3 ]}

\newcommand{\bigO}{\mathcal{O}}

\newcolumntype{C}{>{$}c<{$}}
\newcolumntype{L}{>{$}l<{$}}
\newcolumntype{R}{>{$}r<{$}}

\newenvironment{syntaxleft}
  {\begin{flushleft}\noindent%
   \begin{tabular}{@{}R@{~~}R@{~~}L@{\quad}l}}
  {\end{tabular}\end{flushleft}}

\def\syntaxclass#1{\hss\mbox{#1}\quad\hfill}
\def\syntaxclassn#1{\\[\smallskipamount]\syntaxclass{#1}}

\setcopyright{cc}
\setcctype{by}
\acmDOI{10.1145/3776695}
\acmYear{2026}
\acmJournal{PACMPL}
\acmVolume{10}
\acmNumber{POPL}
\acmArticle{53}
\acmMonth{1}
\received{2025-07-10}
\received[accepted]{2025-11-06}

\begin{document}
\title{A Lazy, Concurrent Convertibility Checker}

\author{Nathanaëlle Courant}
\email{nathanaelle.courant@ocamlpro.com}
\orcid{0000-0002-8736-3060}
\affiliation{%
  \institution{OCamlPro}
  \city{Paris}
  \country{France}
}
\author{Xavier Leroy}
\email{xavier.leroy@college-de-france.fr}
\orcid{0000-0002-8971-9171}
\affiliation{%
  \institution{Collège de France, PSL University}
  \city{Paris}
  \country{France}
}

\begin{abstract}
Convertibility checking --- determining whether two lambda-terms are equal up to reductions --- is a crucial component of proof assistants and dependently-typed languages.  Practical implementations often use heuristics to quickly conclude that two terms are convertible, or are not convertible, without reducing them to normal form.  However, these heuristics can backfire, triggering huge amounts of unnecessary computation.  This paper presents a novel convertibility-checking algorithm that relies crucially on \emph{laziness} and \emph{concurrency}.  Laziness is used to share computations, while concurrency is used to explore multiple convertibility subproblems in parallel or via fair interleaving.  Unlike heuristics-based approaches, our algorithm always finds an easy solution to the convertibility problem, if one exists.  The paper describes the algorithm in process calculus style, discusses its complexity, and reports on its mechanized proof of partial correctness and its lightweight experimental evaluation.
\end{abstract}



\maketitle

\section{Introduction}  \label{s:intro}

Lambda-terms and related functional notations are widely used in functional programming languages, in higher-order type systems, and in higher-order logics.  These terms come with a notion of \emph{reduction}, which expresses elementary steps of computation, such as the famous beta-reduction $(\lam x.t)~u \red t[x \becomes u]$, which expresses the application of a function.

Reductions have two main uses: \emph{evaluation} (reducing a term to a final result) and \emph{conversion} (determining whether two terms are equal up to reductions).  Evaluation accounts for the execution of functional programs and their specialization through partial evaluation techniques.  Conversion is used in type systems and in logics based on type theory to state that two types or two propositions are identical up to computation.  This concept is captured by the well-known typing rule
$$
\inferrule*[right=conv]
  {\Gamma \vdash a : t' \\ t \convertible t'}
  {\Gamma \vdash a : t}
$$
For example, the two propositions $2 + 2 = 4$ and $4 = 4$ are convertible, since $2 + 2$ reduces to $4$; therefore, the trivial proof term $"refl"~4$ (reflexivity of equality) of $4 = 4$ also proves $2 + 2 = 4$; no deduction steps are necessary.  This is an instance of proof by reflection, where deduction and proof search are replaced by computations performed during type checking
\citep{ DBLP:conf/tacs/Boutin97,DBLP:conf/mpc/KokkeS15}.

Due to the \RightTirName{conv} rule, proof assistants and programming languages based on type theory such as Agda, Lean and Rocq verify convertibility of terms at every proof step and type-checking step.  Therefore, it is crucial to find algorithms for convertibility checking that are both correct and efficient.  Many different reduction sequences can be applied to a term; some sequences lead quickly to the desired result, while others can take much longer or diverge.

To evaluate a term, we have \emph{reduction strategies} such as call by name, call by value and call by need. The performance characteristics and efficient implementation of these strategies are well known.  For example, call by need is optimal (in terms of the number of beta-reductions) for weak reduction \cite{DBLP:conf/icfp/Balabonski13}, and high-performance implementations exist.

In contrast, no good reduction strategy is known for checking whether two terms are convertible.  The textbook approach is to reduce both terms to normal form and compare the normal forms for equality.  This approach can perform arbitrary amounts of unnecessary computation (see \secref{s:intuitions} for examples).  Convertibility checkers used by proof assistants perform incremental evaluation of the two terms, bringing them to a state where they are either syntactically equal or obviously non-convertible.  They use heuristics to determine which evaluation to perform next.  As illustrated in \secref{s:intuitions} and~\secref{s:experimental}, these heuristics are sometimes ineffective, performing unnecessary computation and resulting in proofs that take forever to check and proof tactics that take forever to fail.

This paper presents a novel algorithm for checking convertibility that relies crucially on \emph{laziness} and \emph{concurrency}.  Laziness, or more precisely non-strict evaluation, is used to avoid unnecessary computations and to share computations between multiple convertibility subproblems.  Concurrency is used to explore multiple convertibility subproblems in parallel or via fair interleaving, stopping them all as soon as one of them returns conclusive evidence.  Existing convertibility checkers would explore these subproblems sequentially, in an order chosen by heuristics; they can get stuck exploring the wrong subproblem first.  In contrast, our concurrent exploration method will never overlook an easy solution to the convertibility problem, if one exists.

Our convertibility algorithm has been proved sound using the Rocq proof assistant.  It can be easily implemented as an abstract machine.  This machine has the subterm property \cite[\S3.1]{DBLP:conf/rta/AccattoliL12}, meaning that it can be statically compiled to virtual machine code or native code.

The remainder of this paper is organized as follows. 
Section~\ref{s:intuitions} gives examples that illustrate the difficulties of convertibility checking.
Section~\ref{s:laziness} introduces the small process calculus that we use to express our algorithms.
Section~\ref{s:reduction} describes call-by-need evaluation (to WHNF) and normalization.
Sections~\ref{s:convertibility} and~\ref{s:extensions}, the core of the paper, describe the convertibility algorithm.
Its implementation as an abstract machine with explicit scheduling is shown in section~\ref{s:scheduling}, and its Rocq proof of soundness is described in section~\ref{s:rocq-proof}.
Section~\ref{s:performance-analysis} analyses the performance of our algorithm.  Section~\ref{s:experimental} reports on a preliminary experimental evaluation, using the Rocq conversion checker as the baseline.
Related work is discussed in section~\ref{s:related} and followed by concluding remarks in section~\ref{s:conclusions}.

\section{Intuitions on convertibility checking}  \label{s:intuitions}

Consider the problem of determining whether two expressions $e_1$ and $e_2$ are convertible, written $e_1 \convertible e_2$, in the sense that they are equal up to integer arithmetic calculations.  For example, $6 \times 4 + 1 \convertible 10 + 14 + 1$ since $6 \times 4$ is 24 and $10 + 14$ is also 24.

A simple algorithm for determining whether $e_1 \convertible e_2$ is to compute the integer values of~$e_1$ and~$e_2$ and then compare these values for equality.  For example,
\begin{align*}
6 \times 4 + 1 & \convertible 10 + 14 + 1 &&
\text{since $6 \times 4 + 1$ evaluates to 25, as does $10 + 14 + 1$}
\\
6 \times 4 + 1 & \not\convertible 3 \times 8 &&
\text{since $6 \times 4 + 1$ evaluates to 25 and $3 \times 8$ to 24.}
\end{align*}

However, this algorithm can perform unnecessary computations.  For example, let $F$ be an expensive integer function, of cost $\bigO(2^n)$, say.  To determine that
$$ F~20 \convertible F~20 $$
the simple algorithm computes $F~20$ twice.  However, it suffices to note that the two sides of the conversion problem are syntactically identical; therefore, their values must be equal, and no computation is needed.

Similarly, assume that expressions include the lists constructors "cons" and "nil". To determine that
$$ "cons"~(F~20)~"nil" \not\convertible "nil" $$
we do not need to compute $F~20$ at all.  It suffices to observe that the head constructors of the left-hand side ("cons") and of the right-hand side ("nil") are different; therefore, no amount of calculation can make them equal.

Often, the two expressions being tested for convertibility are not identical, but ``fairly close''.  Consider:
$$ F~20 \convertible F~(19 + 1) $$
Again, there is no need to compute both sides.  It suffices to show that $20 \convertible 19 + 1$, by a simple computation.  Then, $F~20 \convertible F~(19 + 1)$ follows immediately.

The previous example suggests the following heuristic: to show that two applications of the same function are convertible, first check if the arguments are pairwise compatible:
$$
F~a_1~\ldots~a_n \convertible F~b_1~\ldots~b_n
\quad \text{if} \quad a_i \convertible b_i \quad \text{for $i = 1, \ldots, n$}
$$
Only when this first check fails should the definition of $F$ be used to further reduce the two sides of the convertibility test.

Unfortunately, this heuristic is not always profitable.  Consider
$$ K~0~(F~20) \convertible K~0~(F~21) $$
where $K$ is the familiar combinator $K~x~y = x$.  
The heuristic above causes $F~20$ and $F~21$ to be computed, which is expensive.  However, if we unroll the definition of $K$ first, the convertibility problem becomes $0 \convertible 0$, which is trivial, and no computation of $F$ is needed.

In many cases, it is preferable to unroll the definition of a recursive function once rather than fully evaluate an application of the function.  For example, let "exp" be the naive exponentiation function
$$ "exp"~n ~=~ "if"~n = 0~"then"~1~"else"~"exp"(n-1) + "exp"(n-1)$$
Consider the convertibility problem
$$ "exp"~40 \convertible "exp"~39 + "exp"~39 $$
Unrolling the definition of "exp" in the left-hand side and simplifying the "if" allows us to prove convertibility without evaluating $"exp"~40$ or $"exp"~39$.

When comparing applications of two different functions, unrolling the definitions of both functions is often the right thing to do, but not always.  Consider the mutually-recursive functions
\begin{align*}
"even"~n & ~=~ "if"~n = 0~"then"~"true"~"else"~"odd"(n-1) \\
"odd"~n & ~=~ "if"~n = 0~"then"~"false"~"else"~"even"(n-1)
\end{align*}
and the convertibility problem
$$ "odd"~999999  \convertible "even"~1000000 $$
The problem can easily be solved by unrolling the definition of "even" in the right-hand side, reducing it to $"odd"~999999$.  However, if we unroll "odd" in the left hand side and "even" in the right-hand side simultaneously, we obtain $"even"~999998 \convertible "odd"~999999$, which is still far from a solution.

As the examples above show, there are many different ways to determine if two expressions are convertible.  Some ways are faster for certain expressions, but no single way is consistently better than the others.  Proof assistants generally use a fixed strategy to determine when and where to perform reductions and unrolling of function definitions.  For instance, given $F~a_1~\cdots~a_n \convertible F~b_1~\cdots~b_n$, the Rocq proof checker first tries to prove $a_i \convertible b_i$ for $i = n, n-1, \ldots, 1$ before unrolling~$F$.  Given $F~a_1~\cdots~a_n \convertible G~b_1~\cdots~b_m$, it chooses whether to unroll $F$ or $G$ based on numerical priorities, which can be controlled with the "Strategy" command \citep{Rocq-9.0}.%
This is a reasonable strategy.  However, any such strategy can go wrong and perform huge amounts of unnecessary computation, which prevent proof checking from completing in a reasonable amount of time \cite[section 2.6.2]{PHD:Gross}.

The approach we propose and develop in this paper is to explore multiple ways to solve a convertibility problem in parallel, instead of trying one way after another based on a fixed strategy.  For instance, when presented with the problem $F~e_1 \convertible F~e_2$, we do not decide whether to begin by solving $e_1 \convertible e_2$, or by unrolling $F$ on the left to obtain $e_1' \convertible F~e_2$, or by unrolling $F$ on the right to obtain $F~e_1 \convertible e_2'$.  Rather, we set up the three corresponding problems and solve them in parallel, stopping them as soon as $e_1 \convertible e_2$ terminates with a ``yes'', or $e_1' \convertible F~e_2$ or $F~e_1 \convertible e_2'$ terminate with a ``yes'' or a ``no''.

In other words, we view solving a convertibility problem as searching a tree of possible proofs of convertibility or non-convertibility.  Using concurrency and fair interleaving, our approach performs a breadth-first traversal of the proof search tree to find the simplest possible proof in this proof space.  In contrast, existing convertibility checkers perform a mostly depth-first traversal of the proof search tree, using strategies to select which branch to explore first. Sometimes, they go down a very long branch and fail to produce a proof in reasonable time.  In the worst case, our breadth-first approach can take time exponential in the length of the shortest proof, but this is still preferable to the existing depth-first approaches, which can take arbitrarily long.

With so many convertibility subproblems being generated and solved in parallel, it is crucial to avoid duplicating computations between subproblems.  For instance, if $G~x = 1 + x$, the problem $G~(F~20) \convertible G~(F~19)$ generates two main subproblems:
$F~20 \convertible F~19$, to compare the arguments to $G$,
and $1 + F~20 \convertible 1 + F~19$, after unrolling $G$ on both sides.
We really want to evaluate $F~20$ and $F~19$ only once, not twice each.  To this end, we systematically use lazy evaluation to share computations within expressions, as in $(\lam x. x + x)~(F~20)$, and between convertibility problems, as in the example above.

Since we are testing the convertibility of lambda-terms, we need a notion of lazy evaluation that extends beyond weak reduction (as in Haskell and other functional languages) to include strong reduction (within the body of a lambda-abstraction).  This is presented in \secref{s:reduction}.  Since we interleave the executions of multiple evaluations and multiple convertibility problems, we need a formulation of lazy evaluation that plays well with concurrency, which is presented in the next section.

\section{Expressing laziness with a process calculus} \label{s:laziness}

The following artificial example illustrates the features we need from the metalanguage used to describe our convertibility testing algorithm.
\begin{eqnarray*}
H(n) & = & "let"~a = F(n)~"and"~b = G(n) ~"in" \\
     &   & "if"~ n < 0 ~"then"~1~"else"~ a \times a \times b
\end{eqnarray*}
We would like to avoid unnecessary evaluations of $F(n)$ and $G(n)$ when evaluating $H(n)$.  Namely:
\begin{itemize}
\item the bindings of $a$ and $b$ should be \emph{lazy}, so that $F(n)$ and $G(n)$ are not evaluated at all if $n < 0$;

\item the computations bound to $a$ and $b$ should be \emph{shared} between multiple uses of these variables, so that $F(n)$ and $G(n)$ are evaluated only once if $n \ge 0$, even though $a$ is used twice;

\item the evaluations of $F(n)$ and $G(n)$ should proceed \emph{in parallel}, or by \emph{fair interleaving}, so that $a \times a \times b$ produces 0 as soon as one of $F(n)$ or $G(n)$ returns 0, without waiting for the other computation to terminate.
\end{itemize}

\begin{figure}
\begin{flushleft}
\begin{syntaxleft}
\syntaxclass{Processes:}
P & ::=  & \sendp{\alpha}{E} & send $E$ on channel $\alpha$ \\
  & \mid & \parp{P_1}{P_2}   & parallel composition \\
  & \mid & \freshp{\alpha}{P} & channel creation
\syntaxclassn{Expressions:}
E & ::=  & x \mid v & variables, values \\
  & \mid & \recvp{E} & receive a value from channel $E$ \\
  & \mid & F~E_1~\cdots~E_n & function applications \\
  & \mid & C~E_1~\cdots~E_n & data constructor applications
\syntaxclassn{Values:}
v & ::=  & \alpha \mid C~v_1~\cdots~v_n & channels, constructors
\end{syntaxleft}%

\medskip

Structural equivalences:
\begin{align*}
\parp{P_1}{P_2} & = \parp{P_2}{P_1} \\
\parp{P_1}{(\parp{P_2}{P_3})} & = \parp{(\parp{P_1}{P_2})}{P_3} \\
\parp{(\freshp{\alpha}{P_1})}{P_2} & =
\freshp{\alpha}{(\parp{P_1}{P_2})}
\quad\mbox{if $\alpha$ not free in $P_2$}
\end{align*}
Generic reduction rule:
$$ \parp{\sendp{\alpha}{v}}{\Gamma[\recvp{\alpha}]} \rightarrow \parp{\sendp{\alpha}{v}}{\Gamma[v]} %
$$
\smallskip
(Plus: specific reduction rules $\sendp{\alpha}{F(\ldots)} \rightarrow \ldots$ for specific functions $F$.)
\end{flushleft}

\caption{The simple process calculus used as a metalanguage in this article.}
\label{f:process-calculus}

\end{figure}

To support these features, we will use a simple process calculus loosely inspired by  the $\pi$-calculus \citep{DBLP:books/daglib/0098267}. %
As summarized in Fig.~\ref{f:process-calculus}, we have \emph{processes} $P$ that execute in parallel ($\parp{P_1}{P_2}$) and communicate values over \emph{channels} ($\alpha, \beta, \gamma, \ldots$).  The process $\sendp{\alpha}{E}$ computes the value of expression~$E$ and sends it over channel $\alpha$.  In an expression, $\recvp{\alpha}$ denotes the value read from channel $\alpha$ when it is available.  Finally, $\freshp{\alpha}{P}$ creates a fresh channel name $\alpha$ for the duration of the execution of $P$.

Using this notation, here is the process that computes $H(n)$ and returns its value on channel $\gamma$:
$$
\sendp{\gamma}{H(n)} ~=~
  \freshp{\alpha,\beta}{~
            \sendp{\alpha}{F(n)}
  \parallel \sendp{\beta}{G(n)}}
  \parallel \sendp{\gamma}{"if"~ n < 0 ~"then"~1~"else"~ \recvp{\alpha} \times \recvp{\alpha} \times \recvp{\beta}}
$$
The two bound variables $a$ and $b$ are represented by two fresh channels $\alpha$ and $\beta$.  Their bindings are represented by the two processes 
$\sendp{\alpha}{F(n)}$ and $\sendp{\beta}{G(n)}$ that run in parallel with the body of $H$.  

The syntax and semantics of the process calculus are summarized in Fig.~\ref{f:process-calculus}.  The crucial part is the value communication rule:
$$ \parp{\sendp{\alpha}{v}}{\Gamma[\recvp{\alpha}]} \red \parp{\sendp{\alpha}{v}}{\Gamma[v]}
$$
It says that if a sending process $\sendp{\alpha}{E}$ has reduced to $\sendp{\alpha}{v}$, where $v$ is the value of $E$, any receiver $\recvp{\alpha}$ in any context $\Gamma$ can be replaced by the value $v$.  Unlike in the $\pi$-calculus, the sending process $\sendp{\alpha}{v}$ remains unchanged, so that all present or future occurrences of $\recvp{\alpha}$ can also be replaced by $v$.

Does the encoding of $H$ in terms of processes satisfy the list of requirements above?
\begin{itemize}
\item \emph{Sharing} of computations bound to variables is enforced by the communication rule.  In $H(n)$ above, $F(n)$ is computed only once, and its value $v$ replaces the two occurrences of $\recvp{\alpha}$.

\item \emph{Parallelism} is inherent in the process calculus encoding of $H$.  The evaluations of $F(n)$ and $G(n)$ can be freely interleaved, and we can give a parallel semantics to the multiplication operator: $0 \times E \red 0$ and $E \times 0 \red 0$, even if $E$ is blocked on a receive operation.

\item \emph{Laziness} is not guaranteed by the process calculus encoding, only \emph{non-strictness}: the evaluations of $F(n)$ and $G(n)$ can start right away, but can also be delayed until the values of $\recvp{\alpha}$ and $\recvp{\beta}$ are required to make progress.  
However, laziness can be enforced by an appropriate \emph{scheduling} of process reductions.  Typically, the processes $\sendp{\alpha}{F(n)}$ and $\sendp{\beta}{G(n)}$ should not be reduced until the values of $\recvp{\alpha}$ and $\recvp{\beta}$ are needed to make progress in the  process that sends on~$\gamma$, that is, until $n$ was tested nonnegative.
\end{itemize}

More generally, we can encode ML-style explicit laziness, presented as two expressions: $"lazy"~E$, which produces a thunk that evaluates $E$ on demand, and $"force"~E$, which forces the thunk $E$ and returns its value.  We represent thunks by channels; thus, $"force"~E$ is simply $\recvp{E}$, and "lazy" has the following reduction rule:
$$
\sendp{\alpha}{\Delta["lazy"~E]} \red
\freshp{\beta}{\parp{\sendp{\alpha}{\Delta[\beta]}}{\sendp{\beta}{E}}}
$$
where $\Delta$ is an expression evaluation context.  Without scheduling restrictions, the evaluation of $E$ can start immediately, making $"lazy"~E$ behave like a future. To enforce laziness, we need to perform only the reductions that are necessary to produce the final value on the result channel $\alpha$.  Consider the following typical intermediate evaluation state:
$$
\sendp{\alpha}{\Delta_0[\recvp{\alpha_1}]} \parallel
\sendp{\alpha_1}{\Delta_1[\recvp{\alpha_2}]} \parallel \cdots \parallel
\sendp{\alpha_{n-1}}{\Delta_{n-1}[\recvp{\alpha_n}]} \parallel
\sendp{\alpha_n}{E} \parallel P
$$
where $E$ can reduce and the $\Delta_i$ are expression evaluation contexts.  There is a chain of computations waiting for values to be sent on channels:  $\alpha$ is waiting for $\alpha_1$, which is waiting for $\alpha_2$, all the way to $\alpha_{n-1}$, which is waiting for $\alpha_n$, which is not waiting on anyone and can make progress (since $E$ can reduce).  Therefore, the reduction of $E$ is the one that must be performed at this point.  Other reductions in the remaining processes $P$ might be possible, but are not required to  progress on the evaluation of $\alpha$, so they are not performed.

\section{Strong call-by-need reduction} \label{s:reduction}

We now use our process calculus notation to describe algorithms that reduce lambda-terms to weak head normal form, then to normal form.  Our algorithms implement call-by-need strategies, using channels and parallel processes to share the reductions of subterms.  However, they make no attempt at sharing subcontexts the way optimal reduction algorithms do.

\subsection{The source language}

While our approach extends to richer functional languages, this paper considers only the pure untyped lambda-calculus, and its extension with defined constants $c$.
\begin{syntaxleft}
\syntaxclass{Pure lambda terms:}
t, u & ::= & x \mid \lambda x. t \mid t~u
\syntaxclassn{Extended lambda terms:}
t, u & ::= & x \mid \lambda x. t \mid t~u \mid c
\end{syntaxleft}%
Defined constants are constants that are bound to terms at top-level.  For example, this is written $"def"~c := t$ in Lean and $"Definition"~c := t.$ in Rocq.  Unlike ordinary constants, which are opaque, defined constants can be expanded (replaced by their definitions) at any time during computation.

\subsection{Reduction to weak head normal form} \label{s:evaluation}

To compute weak head normal forms (WHNF) of pure lambda terms, we use an environment machine inspired by Krivine's machine \cite{DBLP:journals/lisp/Krivine07}, with the main difference that environments $e$ map variables not to unevaluated thunks, but to channels connected to processes that evaluate these thunks.  The machine produces values of the following shape:
\begin{syntaxleft}
\syntaxclass{Values:}
v & ::=  & \shallowclos{x}{t}{e} & closure of function $\lam x.t$ by environment $e$ \\
  & \mid & \neutralv{x}{s} & free variable $x$ applied to arguments $s$
\syntaxclassn{Environments:}
e & ::=  & \{ x_1 \mapsto \alpha_1, \ldots, x_n \mapsto \alpha_n \}
& mapping from variables to channels
\syntaxclassn{Stacks:}
s & ::=  & \alpha_1 \cons \alpha_2 \cdots \alpha_n \cons \nil
& lists of arguments (channels)
\end{syntaxleft}%
To compute the WHNF of term $t$ in environment $e$, the machine starts in state $\evalf{t}{e}$ and performs the following transitions.
\begin{align*}
\sendp{\alpha}{\evalf{t}{e}} & \red
  \sendp{\alpha}{\reducef{t}{e}{\nil}}
\\
\sendp{\alpha}{\reducef{(t\ u)}{e}{s}} &\red
  \freshp{\beta}{\parp
    {\sendp{\alpha}{\reducef{t}{e}{(\beta \cons s)}}}
    {\sendp{\beta}{\evalf{u}{e}}}}
\\
\sendp{\alpha}{\reducef{(\lambda x. t)}{e}{s}} &\red
  \sendp{\alpha}{\applyf{\shallowclos{x}{t}{e}}{s}}
\\
\sendp{\alpha}{\reducef{x}{e}{s}} &\red
  \sendp{\alpha}{\applyf{\recvp{\getenv{e}{x}}}{s}}
\quad\text{if $x \in \Dom(e)$}
\\
\sendp{\alpha}{\reducef{x}{e}{s}} &\red
  \sendp{\alpha}{\applyf{\neutralvar{x}}{s}}
\quad\text{if $x \notin \Dom(e)$}
\\[1mm]
\sendp{\alpha}{\applyf{v}{\nil}} &\red \sendp{\alpha}{v}
\\
\sendp{\alpha}{\applyf{\shallowclos{x}{t}{e}}{(\beta \cons s)}} &\red
  \sendp{\alpha}{\reducef{t}{(\extenv{e}{x}{\beta})}{s}}
\\
\sendp{\alpha}{\applyf{\neutralv{x}{s'}}{s}} &\red
  \sendp{\alpha}{\neutralv{x}{(s' \append s)}}
\end{align*}
In the $\reducef t e s$ state, the machine traverses the spine of applications of $t$, recording arguments on the stack $s$.  If $t$ is an application $t~u$, the machine sets up a new process $\evalf u e$ that reduces~$u$ and sends its value on a fresh channel $\beta$, then pushes $\beta$ on the stack and proceeds with the reduction of $t$.  In all other cases, the machine switches to the $\applyf v s$ state, where $v$ is the value of~$t$: a closure if $t$ is a function abstraction, a neutral value $\neutralvar{x}$ if $t$ is a free variable $x$, and the value read from channel $\getenv{e}{x}$ if $x$ is a bound variable.

In the $\applyf v s$ state, if $s$ is empty, evaluation is finished and $v$ is returned.  If $s$ is not empty and~$v$ is a function closure, a $\beta$-reduction step is performed and the machine resumes in the "reduce" state.  If $v$ is a constant or a free variable already applied to arguments $s'$, the arguments $s$ are added to $s'$.

In terms of reduction strategies, this machine  implements non-strict evaluation with sharing (of the evaluation of a function argument).  The process $\sendp{\beta}{\evalf u e}$ that is created when the application $t~u$ is reduced can start executing immediately or stay idle until the value of a variable $x$ bound to $u$ is needed for the first time.  At that time, the machine computes $\recvp{\getenv e x}$, that is, $\recvp{\beta}$, forcing the process $\sendp{\beta}{\reducef u e \nil}$ to evaluate to $\sendp{\beta}{v}$ for some value $v$.  If the value of $x$ is needed again later, it is obtained from this $\sendp{\beta}{v}$ process; no recomputation is required.  Therefore, depending on the scheduling of processes, the machine implements call-by-value, call-by-need, or any strategy ``in between'', but not call-by-name.  Call-by-need can be obtained by restricting the scheduling of processes appropriately, as outlined at the end of \secref{s:laziness}.

\subsection{Reduction under lambdas and normalization} \label{s:normalization}

Normal forms can be computed by alternating evaluation phases, which produce values (WHNFs), and reification phases, which turn these values into terms in normal forms.  This is similar to the ``eval'' and ``reify'' phases of normalization by evaluation \cite{DBLP:conf/nada/BergerES98} and of type-directed partial evaluation \cite{DBLP:conf/popl/Danvy96}.  For example, if evaluation produces a function closure $\shallowclos{x}{t}{e}$, we can apply it to a fresh free variable $\neutralvar{y}$, obtaining a value $v$, then recursively reify $v$ to a normal-form term $t$, and finally produce the normal form $\lam y. t$.

A naive implementation of this normalization procedure can duplicate evaluations, however.  For example, consider the term $(\lam f. g~f~f)~(\lam x.t)$ where $g$ is a free variable.  Evaluation produces the value $\neutralv{g}{\beta~\beta}$, where $\beta$ is a channel that produces a closure for $\lam x.t$.  Naive reification will reify each occurrence of this closure independently, causing $t$ to be normalized twice.

As described in \cite{DBLP:journals/pacmpl/BiernackaCD22} and independently observed by us, this unsharing of function values can be avoided by anticipating the need to reduce in the function body $t$ when creating the closure for the function $\lam x. t$.  In our channel-based presentation, this means adding two components to every function closure $\shallowclos{x}{t}{e}$: a fresh free variable $y$ used for normalization, and a channel $\delta$ connected to a process that evaluates on demand the application of $t$ to $y$.
\begin{syntaxleft}
\syntaxclass{Values:}
v & ::=  & \deepclos{x}{t}{e}{y}{\delta} & enriched function closure \\
  & \mid & \neutralv{x}{s}
\end{syntaxleft}%
The evaluation rule for function abstractions becomes:
\begin{align*}
\sendp{\alpha}{\reducef{(\lambda x. t)}{e}{s}} &\red 
 \freshp{\gamma \delta}{\parp{\parp
   {\sendp{\alpha}{\applyf{\deepclos{x}{t}{e}{y}{\delta}}{s}}}
   {\sendp{\delta}{\evalf{t}{(\extenv{e}{x}{\gamma})}}}}
   {\sendp{\gamma}{\neutralvar{y}}}} \\
& \qquad \qquad \text{where $y$ is a fresh variable}
\intertext{The extra components of closures are ignored during application:}
\sendp{\alpha}{\applyf{\deepclos{x}{t}{e}{y}{\delta}}{(\beta \cons s)}} &\red
  \sendp{\alpha}{\reducef{t}{(\extenv{e}{x}{\beta})}{s}}
\end{align*}
With this twist on closures, we can define $\normalizef t$, the normalization of a closed term $t$, and $\readbackf v$, the reification of a value $v$ as a lambda-term, by the following rules:
\begin{align*}
\sendp{\alpha}{\normalizef{t}} & \red
  \freshp{\beta}{
    {\sendp{\alpha}{\readbackf {\recvp{\beta}}}}
  \parallel
    {\sendp{\beta}{\evalf{t}{\emptyenv}}}}
\\
\readbackf{\deepclos{x}{t}{e}{y}{\delta}} & \red
  \lam y. \readbackf{\recvp{\delta}}
\\
\readbackf{\neutralv{x}{\beta_1 \cdots \beta_n}} & \red
  x~(\readbackf{\recvp{\beta_1}})~\cdots~(\readbackf{\recvp{\beta_n}})
\end{align*}

\subsection{Defined constants} \label{s:defined-constants}

We now extend the evaluation and reification approach of sections~\ref{s:evaluation} and~\ref{s:normalization} to defined constants.  Unlike "let"-bound variables, which have local scope and must therefore be expanded during reduction to WHNF, defined constants have global scope and can remain unexpanded in WHNFs, resulting in values $\neutralv{c}{s}$ that carry an unexpanded constant $c$ and a possibly empty list of arguments $s$ to which $c$ is applied.  This enables faster convertibility testing of two terms,  as outlined in \secref{s:intuitions} and developed in \secref{s:convertibility}.  However, when unfolding a defined constant $c$ to reduce a value $\neutralv{c}{s}$, we must be careful not to duplicate the evaluation of the definition $t$ of $c$ or its application to the arguments $s$.  To this end, we reuse the approach of \secref{s:normalization}: we attach a channel $\delta$ to each $\neutralv{c}{s}$ value and connect this channel to a process that evaluates this application of $c$ on demand.
\begin{syntaxleft}
\syntaxclass{Values:}
v & ::=  & \deepclos{x}{t}{e}{y}{\delta} \\
  & \mid & \neutralv{x}{s} \\
  & \mid & \constv{c}{s}{\delta} & constant $c$ applied to $s$, with actual value available from $\delta$
\end{syntaxleft}%

Given a set of constant definitions $c_1 := t_1, \ldots, c_n := t_n$, we define a global environment $K$ that maps constants to fresh channels, and a process $KP$ that evaluates the $t_i$ on demand and sends their values to those channels.
\begin{align*}
K & = \{ c_1 \mapsto \alpha_1; \ldots; c_n \mapsto \alpha_n \}
\\
KP & = \sendp{\alpha_1}{\evalf{t_1}{\emptyenv}}
       \parallel \cdots \parallel
       \sendp{\alpha_n}{\evalf{t_n}{\emptyenv}}
\end{align*}
When we evaluate a constant $c$, we pair it with the channel $K(c)$, thus ensuring that the evaluation of its definition $t$ is properly shared and non-strict:
\begin{align*}
\sendp{\alpha}{\reducef{c}{e}{s}} &\red
  \sendp{\alpha}{\applyf{(\constvinit{c}{K(c)})}{s}}
\intertext{New channels and new processes are created when a constant is applied to new arguments:}
\sendp{\alpha}{\applyf{(\constv{c}{s'}{\delta})}{s}} &\red
  \freshp{\gamma}{\parp
     {\sendp{\alpha}{\constv{c}{(s' \append s)}{\gamma}}}
     {\sendp{\gamma}{\applyf{\recvp{\delta}}{s}}}}
\intertext{The normalization procedure of \secref{s:normalization} is modified as follows:}
\sendp{\alpha}{\normalizef{t}} & \red
  \freshp{\beta}{
    {\sendp{\alpha}{\readbackf {\recvp{\beta}}}}
  \parallel
    {\sendp{\beta}{\evalf{t}{\emptyenv}}}
  \parallel
    KP}
\\
\readbackf{\deepclos{x}{t}{e}{y}{\delta}} & \red
  \lam y. \readbackf{\recvp{\delta}}
\\
\readbackf{\neutralv{x}{\beta_1 \cdots \beta_n}} & \red
  x~(\readbackf{\recvp{\beta_1}})~\cdots~(\readbackf{\recvp{\beta_n}})
\\
\readbackf{(\constv{c}{s}{\delta})} &\red
  \readbackf{\recvp{\delta}}
\end{align*}

\section{Convertibility testing}  \label{s:convertibility}

\subsection{The basic algorithm} \label{s:basic-convertibility}

Just like normalization can be viewed as a combination of evaluation and reification, determining whether two terms $t, t'$ are convertible can be viewed as a combination of evaluation of~$t$ and~$t'$ and comparison of the resulting values~$v, v'$.  For example, if $v$ and $v'$ are applications of free variables $x$ and $x'$, we need to check that $x = x'$ and that the arguments are pairwise convertible.

More precisely, to test the convertibility of $t$ and $t'$ and send the resulting Boolean value to channel $\alpha$, we start with the following process:
$$
\freshp{\beta\beta'}{
    \sendp{\alpha}{\convf{\beta}{\beta'}{\nil}}
  \parallel
    \sendp{\beta}{\evalf{t}{\emptyenv}}
  \parallel
    \sendp{\beta'}{\evalf{t'}{\emptyenv}}}
  \parallel
    KP
$$
Here, "eval" is the reduction to WHNF from \secref{s:reduction}, and $\convf{\beta}{\beta'}{\xi}$ is the comparison of the values read from the channels $\beta$ and $\beta'$ up to a renaming $\xi$ of free variables (a list of pairs of variables that are considered equal).  The function $\convfname$ and the auxiliary functions $\convvfname$ and $\cstkfname$ are defined by the following rules:
\begin{align*}
\sendp{\alpha}{\convf{\beta}{\beta'}{\xi}} & \red
\sendp{\alpha}{\convvf{\recvp{\beta}}{\recvp{\beta'}}{\xi}}
\\[1mm]
\sendp{\alpha}{\convvf{\deepclos{x}{t}{e}{y}{\delta}}
                      {\deepclos{x'}{t'}{e'}{y'}{\delta'}}
                      {\xi}} & \red
\sendp{\alpha}{\convf{\delta}{\delta'}{((y, y') \cons \xi)}}
\\
\sendp{\alpha}{\convvf{\neutralv{x}{s}}
                     {\neutralv{x'}{s'}}
                     {\xi}} &\red
    \sendp{\alpha}{\cstkf{s}{s'}{\xi}}
      \quad \text{if $(x, x') \in \xi$ and $\listlength{s} = \listlength{s'}$} \\
\sendp{\alpha}{\convvf{v_1}{v_2}{\xi}} &\red
  \sendp{\alpha}{\falseb} \quad \text{in all other cases}
\\[1mm]
\sendp{\alpha}{\cstkf{\beta_1 \cdots \beta_n}{\beta'_1 \cdots \beta'_n}{\xi}}
& \red
\freshp{\gamma_1 \ldots \gamma_n}{
  \begin{array}[t]{@{~~}l@{}}
    \sendp{\alpha}{\recvp{\gamma_1} \land \ldots \land \recvp{\gamma_n}} \\
  \parallel
    \sendp{\gamma_1}{\convf{\beta_1}{\beta'_1}{\xi}}
  \parallel \cdots \parallel
    \sendp{\gamma_n}{\convf{\beta_n}{\beta'_n}{\xi}}
  \end{array}
}
\end{align*}
(We have only shown the cases for pure lambda terms.  The cases for defined constants $c$ are discussed in \secref{s:convertibility-constants}.)

If the values read from channels $\beta$ and $\beta'$ are two closures $\deepclos{x}{t}{e}{y}{\delta}$ and $\deepclos{x'}{t'}{e'}{y'}{\delta'}$, we recursively compare the WHNFs of the corresponding function bodies, which can be read from $\delta$ and $\delta'$, up to equality of the variables $y$ and $y'$, which we express by adding the pair $(y, y')$ to the current renaming $\xi$.  (The freshness requirements on the variables stored in extended function closures guarantee that $y$ and $y'$ are not already involved in the renaming $\xi$, as shown in the Rocq proof described in \secref{s:rocq-proof}.)

If the values read from channels $\beta$ and $\beta'$ are two applications of free variables $\neutralv{x}{s}$ and $\neutralv{x'}{s'}$, we check that the variables $x$ and $x'$ are equal up to the current renaming $\xi$, that the argument lists $s$ and $s'$ have the same length, and that the arguments are pairwise convertible, as expressed by $\cstkf {s}{s'}{\xi}$.  We could have written
$$
\sendp{\alpha}{\convvf{\neutralv{x}{\beta_1 \cdots \beta_n}}
                     {\neutralv{x'}{\beta'_1 \cdots \beta'_n}}
                     {\xi}} \red
\sendp{\alpha}{\convf{\beta_1}{\beta'_1}{\xi} \land \ldots \land
               \convf{\beta_n}{\beta'_n}{\xi}}
\quad \text{if $(x, x') \in \xi$}
$$
The more convoluted definition above, using auxiliary processes and fresh channels, makes it obvious that the sub-convertibility tests $\convf{\beta_i}{\beta'_i}{\xi}$ can run in parallel.

In the definition of $\cstkfname$, the $\land$ operator is ``parallel and'': it reduces to false as soon as one of its arguments reduces to false, even if the other argument is blocked reading from a channel.
$$  \trueb \land \trueb \red \trueb \qquad
    \falseb \land E \red \falseb \qquad
    E \land \falseb \red \falseb \qquad
$$
Combined with the non-strictness of "eval" (reduction to WHNF), this gives our convertibility test nice early-failure properties.  For example, if $t_1$ and $t_2$ are terms that are expensive to compute, and $x_1$ and $x_2$ are two different free variables, the test determines that $x_1~t_1 \not\convertible x_2~t_2$ without ever computing $t_1$ nor $t_2$: the two terms reduce to the values $v_1 = \neutralv{x_1}{\beta_1}$ and $v_2 = \neutralv{x_2}{\beta_2}$, where~$\beta_1$ and~$\beta_2$ are channels connected to processes that evaluate $t_1$ and $t_2$, and the comparison $\convvf{v_1}{v_2}{\xi}$ returns $\falseb$ immediately, since $(x_1, x_2) \notin \xi$.

Likewise, the test can determine that $x~x_1~t_1 \not\convertible x~x_2~t_2$ without computing $t_1$ or $t_2$ in full. Two convertibility subproblems are generated, one corresponding to $x_1 \convertible x_2$ and the other to $t_1 \convertible t_2$:
$$
\sendp{\alpha}{\recvp{\gamma_1} \land \recvp{\gamma_2}}
\parallel
\sendp{\gamma_1}{\convvf{\neutralvar{x_1}}{\neutralvar{x_2}}{\xi}}
\parallel
\sendp{\gamma_2}{\convf{\beta_1}{\beta_2}{\xi}}
\parallel \cdots
$$
Assuming fair interleaving, $\convvf{\neutralvar{x_1}}{\neutralvar{x_2}}{\xi}$ quickly reduces to $\falseb$, causing $\falseb$ to be sent on $\alpha$, while $\convf{\beta_1}{\beta_2}{\xi}$ has barely started to evaluate $t_1$ and $t_2$

Since reduction to WHNF preserves sharing, our convertibility test can also avoid some repeated evaluations that a more naive algorithm would perform.  For example, to determine that $(\lam x. f~x~x)~t \convertible (\lam y. f~y~y)~t$, where $f$ is a free variable, it evaluates $t$ only twice, once for the left-hand side occurrence of $t$ and once for the right-hand side occurrence.  (Section~\ref{s:sharing-convertibility-processes} shows one way to also share the convertibility processes, not just the evaluation processes.  Section \ref{s:sharing-subterms} shows one way to avoid evaluating $t$ at all.)

\subsection{Handling defined constants} \label{s:convertibility-constants}

We now extend the basic convertibility test from \secref{s:basic-convertibility} to support defined constants as introduced in \secref{s:defined-constants}.  As the examples in \secref{s:intuitions} demonstrate, there is no optimal strategy to handle defined constants in a convertibility test: in general, constants have to be unfolded (replaced by their definitions) to determine convertibility; but in some cases, the test can conclude that two terms are convertible without unfolding constants, treating them as simple names instead, which can avoid unnecessary computations.  Our algorithm strives to keep all possibilities open, exploring them in parallel and relying on non-strict evaluation and early-failure and early-success optimizations to shorten this exploration.

Consider again the comparison $\convvf{v}{v'}{\xi}$ of two values.  If one value is a possibly applied, defined constant $\constv{c}{s}{\delta}$ and the other is a different kind of value, there is no choice but to unfold the definition of $c$ and continue reducing to WHNF.  The resulting value can simply be read from $\delta$, since the evaluation that produced $\constv{c}{s}{\delta}$ anticipated this need (as explained in \secref{s:defined-constants}).
\begin{align*}
  \sendp{\alpha}{\convvf{\constv{c}{s}{\delta}}{v}{\xi}} &\red
     \sendp{\alpha}{\convvf{\recvp{\delta}}{v}{\xi}} 
     \qquad \text{if $v$ is not a constant}
\\
  \sendp{\alpha}{\convvf{v}{\constv{c}{s}{\delta}}{\xi}} &\red
     \sendp{\alpha}{\convvf{v}{\recvp{\delta}}{\xi}}
     \qquad \text{if $v$ is not a constant}
\end{align*}

If the values $v$ and $v'$ are the same constant, or more generally the same application of a constant, that is, $\constv{c}{s}{\delta}$ and $\constv{c'}{s'}{\delta'}$ with $\delta = \delta'$, we know that $c = c'$ and $s = s'$ and we can return $\trueb$ immediately.
\begin{align*}
\sendp{\alpha}{\convvf{\constv{c}{s}{\delta}}{\constv{c'}{s'}{\delta}}{\xi}}
&\red \trueb
\end{align*}

If the values $v$ and $v'$ are applications of different defined constants $\constv{c}{s}{\delta}$ and $\constv{c'}{s'}{\delta'}$, with $c \not= c'$, it is tempting to unfold both $c$ and $c'$ in one step.  However, as exemplified in \secref{s:intuitions}, this can result in missed opportunities to conclude quickly that the two values are convertible.  Instead, we explore the two possibilities (unfold $c$ or unfold $c'$) in parallel, and choose whichever result is obtained first.
\begin{align*}
\sendp{\alpha}{\convvf{\constv{c}{s}{\delta}}{\constv{c'}{s'}{\delta'}}{\xi}}
&\red
\freshp{\beta\gamma}{
\begin{array}[t]{@{~}l}
    \sendp{\alpha}{\recvp{\beta} \choice \recvp{\gamma}}
\\ \parallel
    \sendp{\beta}{\convvf{\constv{c}{s}{\delta}}{\recvp{\delta'}}{\xi}}
\\ \parallel
    \sendp{\gamma}{\convvf{\recvp{\delta}}{\constv{c'}{s'}{\delta'}}{\xi}}
\end{array}
}
\quad\text{if $c \neq c'$ or $\listlength{s} \not= \listlength{s'}$}
\end{align*}
The Boolean choice operator $\choice$ returns whichever of its two arguments terminates first, knowing that they evaluate to the same Boolean value.  (Unfolding a constant in the left-hand side or in the right-hand side doesn't change the Boolean value of the convertibility test.)
$$  \trueb \choice E \red \trueb \qquad
    \falseb \choice E \red \falseb \qquad
    E \choice \trueb \red \trueb \qquad
    E \choice \falseb \red \falseb
$$

Finally, if the two values being compared are applications $\constv{c}{s}{\delta}$ and $\constv{c}{s'}{\delta'}$ of the same defined constant $c$ to the same number of arguments, a third possibility arises: just compare the arguments pairwise and return $\trueb$ if they are pairwise convertible, as in the case of free variables or abstract constants.
\begin{align*}
  \sendp{\alpha}{\convvf{\constv{c}{s}{\delta}}{\constv{c}{s'}{\delta'}}{\xi}}
&\red
\freshp{\beta\gamma\eta}{
\begin{array}[t]{@{~}l}
    \sendp{\alpha}{\recvp{\eta} \biasedor (\recvp{\beta} \choice \recvp{\gamma})}
\\ \parallel
    \sendp{\beta}{\convvf{\constv{c}{s}{\delta}}{\recvp{\delta'}}{\xi}}
\\ \parallel
    \sendp{\gamma}{\convvf{\recvp{\delta}}{\constv{c'}{s'}{\delta'}}{\xi}}
\\
   \parallel
    \sendp{\eta}{\cstkf{s}{s'}{\xi}}
\end{array}
}
\quad\text{if $\listlength{s} = \listlength{s'}$}
\end{align*}
The ``biased choice'' operator $E_1 \biasedor E_2$ returns the Boolean value of $E_2$ under the assumption that $E_1$ implies $E_2$.  Therefore, if $E_1$ terminates early with value $\trueb$, $E_1 \biasedor E_2$ can return $\trueb$ without waiting for $E_2$ to terminate; and if $E_2$ terminates early, its value can be returned immediately, without waiting for $E_1$ to terminate.
$$  \trueb \biasedor E \red \trueb \qquad
    E \biasedor \trueb \red \trueb \qquad
    E \biasedor \falseb \red \falseb
$$
Here, we use a combination $E_1 \biasedor (E_2 \choice E_3)$ of biased choice and regular choice, where $E_1$ is ``the argument lists $s$ and $s'$ are pairwise convertible'', $E_2$ is ``$c~s \convertible c~s'$ after unrolling $c$ in the left-hand side'', and $E_3$ is ``$c~s \convertible c~s'$ after unrolling $c$ in the right-hand side''.  As soon as one of $E_1$, $E_2$ or $E_3$ returns $\trueb$, we know that $c~s$ and $c~s'$ are convertible and can immediately return $\trueb$.  As soon as one of $E_2$ or $E_3$ returns $\falseb$, we know that $c~s$ and $c~s'$ are not convertible and can return $\falseb$.  If $E_1$ returns~$\falseb$, we know that the arguments are not convertible but cannot conclude anything about the convertibility of $c~s$ and $c~s'$.
(Consider $c = \lam x. \lam y. x$ and $s, s'$ differing in their second elements.)

\section{Extensions}  \label{s:extensions}

\subsection{Avoiding redundant unfolding of constants}
\label{s:no-unfold-constants}

The progressive unfolding of defined constants, on either the left-hand side or the right-hand side but not both sides simultaneously, naturally leads to duplicated $\convvfname$ tests.  For example, if $c \not= c'$ are defined constants, and assuming that $\recvp{\delta}$ reduces to $v$ and $\recvp{\delta'}$ to $v'$,
\begin{align*}
\sendp{\alpha}{\convvf{\constvinit{c}{\delta}}{\constvinit{c'}{\delta'}}{\xi}}
& \redplus
\freshp{\beta\gamma}{
    \sendp{\alpha}{\recvp{\beta} \choice \recvp{\gamma}}
 \parallel
    \sendp{\beta}{\convvf{\constvinit{c}{\delta}}{v'}{\xi}}
 \parallel
    \sendp{\gamma}{\convvf{v}{\constvinit{c'}{\delta'}}{\xi}}
}
\\
& \redplus
\freshp{\beta\gamma}{
    \sendp{\alpha}{\recvp{\beta} \choice \recvp{\gamma}}
 \parallel
    \sendp{\beta}{\convvf{v}{v'}{\xi}}
 \parallel
    \sendp{\gamma}{\convvf{v}{v'}{\xi}}
}
\end{align*}
We have two identical processes $\convvf{v}{v'}$ that run concurrently.   Evaluations within~$v$ and~$v'$ will be shared, and one processes will become unneeded as soon as the other terminates.  Nonetheless, some convertibility testing work is duplicated, and the number of convertibility processes can increase exponentially.

To avoid that, we introduce \emph{applied frozen constants} as a new type of values:
\begin{syntaxleft}
\syntaxclass{Values:}
v & ::=  & \deepclos{x}{t}{e}{y}{\delta}
   \mid \neutralv{x}{s}
   \mid \constv{c}{s}{\delta} \\
  & \mid &\frozenconstv{c}{s} & frozen constant $c$ applied to $s$
\end{syntaxleft}%
Frozen constants cannot be unfolded.  During evaluation, they behave like free variables:
$$
\sendp{\alpha}{\applyf{\frozenconstv{c}{s}}{s'}} \red
  \sendp{\alpha}{\frozenconstv{c}{(s \append s')}}
$$
Consequently, the only way for a value $v$ to be convertible with a frozen constant $\frozenconstv{c}{s}$ is for $v$ to reduce (possibly by unfolding) to the same constant $c$ applied to some arguments $s'$, with the arguments $s$ and $s'$ being pairwise convertible.

When comparing two applied defined constants $\constv{c}{s}{\delta}$ and  $\constv{c'}{s'}{\delta'}$, we still create two parallel processes, one that unfolds~$c$ and another that unfolds~$c'$.  However, in the process that unfolds~$c'$, we freeze~$c$, replacing $\constv{c}{s}{\delta}$ with $\frozenconstv{c}{s}$.  This way, further unfoldings of $c$ will only take place in one of the processes, but not in both.
\begin{align*}
\sendp{\alpha}{\convvf{\constv{c}{s}{\delta}}{\constv{c'}{s'}{\delta'}}{\xi}}
&\red
\freshp{\beta\gamma}{
\begin{array}[t]{@{~}l}
    \sendp{\alpha}{\recvp{\beta} \biasedor \recvp{\gamma}}
\\ \parallel
    \sendp{\beta}{\convvf{\frozenconstv{c}{s}}{\recvp{\delta'}}{\xi}}
\\ \parallel
    \sendp{\gamma}{\convvf{\recvp{\delta}}{\constv{c'}{s'}{\delta'}}{\xi}}
\end{array}
}
\quad\text{if $c \neq c'$ or $\listlength{s} \not= \listlength{s'}$}
\\
  \sendp{\alpha}{\convvf{\constv{c}{s}{\delta}}{\constv{c}{s'}{\delta'}}{\xi}}
&\red
\freshp{\beta\gamma\eta}{
\begin{array}[t]{@{~}l}
    \sendp{\alpha}{\recvp{\eta} \biasedor (\recvp{\beta} \biasedor \recvp{\gamma})}
\\ \parallel
    \sendp{\beta}{\convvf{\frozenconstv{c}{s}}{\recvp{\delta'}}{\xi}}
\\ \parallel
    \sendp{\gamma}{\convvf{\recvp{\delta}}{\constv{c'}{s'}{\delta'}}{\xi}}
\\
   \parallel
    \sendp{\eta}{\cstkf{s}{s'}{\xi}}
\end{array}
}
\quad\text{if $\listlength{s} = \listlength{s'}$}
\end{align*}
Note the use of a biased choice in $\recvp{\beta} \biasedor \recvp{\gamma}$: the $\beta$ process, which is the one that freezes~$c$, may return~$\falseb$ even though the two values are convertible; only the~$\gamma$ process returns an authoritative result.

When they appear as arguments in a convertibility test, applications of frozen constants $\frozenconstv{c}{s}$ are treated almost like applications of free variables $\neutralv{x}{s}$.  However, a special case is needed when comparing an application of a frozen constant with an application of the same constant that is not frozen.
\begin{align*}
\sendp{\alpha}{\convvf{\frozenconstv{c}{s}}
                     {\frozenconstv{c}{s'}}
                     {\xi}} &\red
    \sendp{\alpha}{\cstkf{s}{s'}{\xi}}
      \qquad \text{if $\listlength{s} = \listlength{s'}$} 
\\
\sendp{\alpha}{\convvf{\frozenconstv{c}{s}}
                      {\constv{c}{s'}{\delta'}}
                      {\xi}} &\red
\freshp{\beta\gamma}{
\begin{array}[t]{@{~}l}
  \sendp{\alpha}{\recvp{\beta} \vee \recvp{\gamma}}
\\
  \parallel \sendp{\beta}{\cstkf{s}{s'}{\xi}}
\\
  \parallel \sendp{\gamma}{\convvf{\frozenconstv{c}{s}}{\recvp{\delta'}}{\xi}}
\end{array}
}
      \quad \text{if $\listlength{s} = \listlength{s'}$} 
\\
\sendp{\alpha}{\convvf{\constv{c}{s}{\delta}}
                      {\frozenconstv{c}{s'}}
                      {\xi}} &\red
\freshp{\beta\gamma}{
\begin{array}[t]{@{~}l}
  \sendp{\alpha}{\recvp{\beta} \vee \recvp{\gamma}}
\\
  \parallel \sendp{\beta}{\cstkf{s}{s'}{\xi}}
\\
  \parallel \sendp{\gamma}{\convvf{\recvp{\delta}}{\frozenconstv{c'}{s}}{\xi}}
\end{array}
}
      \quad \text{if $\listlength{s} = \listlength{s'}$} 
\end{align*}

\subsection[Handling eta-conversion]{Handling $\eta$-conversion} \label{s:eta-conversion}

Several proof assistants consider terms equal up to $\eta$-conversion of functions: $\lam x. M~x \convertible M$.  Our algorithm can be extended to handle $\eta$-conversion as well. The key observation is that, in the presence of $\eta$-conversion, in order to prove $\lam x. M \convertible N$, it suffices to show $M[x \becomes y] \convertible N~y$ where~$y$ is a fresh variable.  However, this should only be attempted when it is safe and profitable to do so.  For example, if $N$ is a lambda-abstraction, this approach is not profitable; and if $N$ is a pair $(N_1, N_2)$ (in an extension of our lambda-calculus with pairs), this approach creates a term $(N_1, N_2)~y$ that goes wrong during evaluation.

The first safe and profitable case is the comparison of a function value with a neutral value, that is, an applied free variable $\neutralv{z}{s}$ or an applied frozen constant $\frozenconstv{c}{s}$.  Then, we can safely apply the neutral value to a fresh variable.
\begin{align*}
\sendp{\alpha}{\convvf{\deepclos{x}{t}{e}{y}{\delta}}
                      {\neutralv{z}{s}}
                      {\xi}} & \red
\freshp{\beta}{
    \sendp{\alpha}{\convvf{\recvp{\delta}}{\neutralv{z}{(s \cons \beta)}}{((y, y') \cons \xi)}} \parallel \sendp{\beta}{\neutralvar{y'}}
} \quad \text{$y'$ fresh}
\end{align*}
(We omit three similar rules: one with $\frozenconstv{c}{s}$ instead of $\neutralv{z}{s}$ and two with the function value and the neutral value swapped.)

When one side is an abstraction and the other is a defined constant, we explore two possibilities in parallel: unfolding the constant or applying it to a fresh free variable.  In the latter case, we must prevent further unfolding of the constant by using the frozen constant mechanism of \secref{s:no-unfold-constants}.  Unfolding the constant after applying it to a fresh free variable can be unsafe (\emph{e.g.}~if the constant expands to a pair) and is not profitable.  The corresponding rule is as follows:
\begin{align*}
\sendp{\alpha}{\convvf{\deepclos{x}{t}{e}{y}{\delta}}
                      {\constv{c}{s}{\delta'}}
                      {\xi}} & \red
\freshp{\beta\gamma\eta}{
\begin{array}[t]{@{~}l}
    \sendp{\alpha}{\recvp{\eta} \biasedor \recvp{\gamma}}
\\ \parallel
    \sendp{\eta}{\convvf{\recvp{\delta}}{\frozenconstv{c}{(s \cons \beta)}}{((y, y') \cons \xi)}}
\\ \parallel
    \sendp{\gamma}{\convvf{\deepclos{x}{t}{e}{y}{\delta}}{\recvp{\delta'}}{\xi}}
\\
   \parallel
    \sendp{\beta}{\neutralvar{y'}}
\end{array}
}\quad\text{$y'$ fresh}
\end{align*}
(Plus a similar rule for
$\convvf{\constv{c}{s}{\delta'}} {\deepclos{x}{t}{e}{y}{\delta}} {\xi}$.)

\subsection{Sharing convertibility processes}
\label{s:sharing-convertibility-processes}

As mentioned at the end of \secref{s:basic-convertibility}, our convertibility checker benefits from our lazy evaluator's ability to share the repeated evaluations of sub-terms.  However, the convertibility tests themselves are not shared and can easily be duplicated.

Consider the test $(\lam x. f~x~x)~t \convertible (\lam y. f~y~y)~u$ where $f$ is a free variable.  After the initial evaluation steps, we end up comparing two inert values $\neutralv{f}{\beta~\beta}$ and $\neutralv{f}{\gamma~\gamma}$:
$$
\sendp{\alpha}{\convvf{\neutralv{f}{\beta~\beta}}{\neutralv{f}{\gamma~\gamma}}{\xi}}
\parallel \sendp{\beta}{\evalf{t}{\emptyenv}}
\parallel \sendp{\gamma}{\evalf{u}{\emptyenv}}
$$
This process further reduces to
$$
\sendp{\alpha}{\recvp{\delta} \land \recvp{\eta}} 
\parallel \sendp{\delta}{\convf{\beta}{\gamma}{\xi}}
\parallel \sendp{\eta}{\convf{\beta}{\gamma}{\xi}}
\parallel \sendp{\beta}{\evalf{t}{\emptyenv}}
\parallel \sendp{\gamma}{\evalf{u}{\emptyenv}}
$$
The evaluations of $t$ and $u$ remain shared, but the convertibility test $\convf{\beta}{\gamma}{\xi}$ is duplicated.  In some cases involving recursive functions, this can result in exponential duplication of convertibility tests (see the "perfect" example in \secref{s:experimental}).

This problem could be avoided by re-sharing $\convf{\beta}{\gamma}{\xi}$ processes as they are created, using hash-consing on the channels $\beta$ and $\gamma$.  On the example above, this re-sharing would result in
$$
\sendp{\alpha}{\recvp{\delta} \land \recvp{\delta}}
\parallel \sendp{\delta}{\convf{\beta}{\gamma}{\xi}}
\parallel \sendp{\beta}{\evalf{t}{\emptyenv}}
\parallel \sendp{\gamma}{\evalf{u}{\emptyenv}}
$$
with a single $\convf{\beta}{\gamma}{\xi}$ process whose output is used twice.

Re-sharing convertibility processes also solves the issue with redundant unfolding of constants described in \secref{s:no-unfold-constants}, without the need to introduce frozen constant values and treat them specially.  However, frozen constant values have other uses, \emph{e.g.}\ to handle $\eta$-conversion, and they are cheaper to implement than process re-sharing.

\subsection{Sharing identical source subterms} \label{s:sharing-subterms}

Evaluation processes $\evalf{t}{e}$ carefully avoid duplicating computations when performing a beta-reduction or unfolding a constant.  Consequently, $(\lam x. x + x)~t \convertible 0$ evaluates $t$ only once, and $1 + c \convertible c + 1$ evaluates the definition of $c$ only once.  However, multiple occurrences of the same source subterm $t$ are evaluated independently.  For example, $t + t \convertible 0$ evaluates $t$ twice, and so does $1 + t \convertible t + 1$.

To share more evaluations, we can "let"-bind some subterms of the source terms before starting the convertibility test.  For example, $t + t$ can be replaced by $"let"~x = t~"in"~x+x$.  To share subterms that occur on both sides of a convertibility test, we need "let" bindings that cover both sides, leading to  generalized convertibility problems of the following form:
$$
"let"~x_1 = t_1~"in"~\ldots~"let"~x_n = t_n~"in"~t \convertible t'
$$
For example, $1 + t \convertible t + 1$ can be replaced by $"let"~x = t~"in"~1 + x \convertible x + 1$.  More generally,
$C[t] \convertible C'[t]$
can be replaced by
$"let"~x = t~"in"~C[x] \convertible C'[x]$
where $x$ is fresh, provided that $t$ does not depend on variables that are bound by $C$ or $C'$.  This transformation includes as a special case the lifting of maximal free expressions described by \citet[chapter 15]{DBLP:books/ph/Jones87} to implement full laziness \citep{DBLP:conf/popl/Balabonski12}.

One way to implement the transformation outlined above is to perform hash-consing on the closed subterms of the original convertibility problem $t \convertible t'$.  This yields (in linear time) a DAG where multiple occurrences of the same closed subterm are shared.  Then, we transcribe (in linear time) this DAG as a set of "let" bindings to obtain a generalized convertibility problem of the form shown above.

To evaluate this generalized convertibility problem, we set up the following process:
$$
\freshp{\beta \beta' \gamma_1 \ldots \gamma_n}{
\begin{array}[t]{@{~}l}
    \sendp{\alpha}{\convf{\beta}{\beta'}{\nil}}
\\
\parallel
    \sendp{\beta}{\evalf{t}{\{x_1 \mapsto \gamma_1, \ldots, x_n \mapsto \gamma_n\}}}
\parallel
    \sendp{\beta'}{\evalf{t'}{\{x_1 \mapsto \gamma_1, \ldots, x_n \mapsto \gamma_n\}}}
\\
\parallel \sendp{\gamma_1}{\evalf{t_1}{\emptyenv}}
\parallel \ldots
\parallel \sendp{\gamma_n}{\evalf{t_n}{\{x_1 \mapsto \gamma_1, \ldots, x_{n-1} \mapsto \gamma_{n-1}\}}}
\end{array}
}
$$
Since it is now possible for the same channel to appear on both sides of a convertibility sub-problem, we add a new early-success case:
$$
 \sendp{\alpha}{\convf{\beta}{\beta'}{\xi}} \red \trueb
\quad \text{if $\beta = \beta'$}
$$
For example, $c~t \convertible c~t$, encoded as $"let"~x=t~"in"~c~x \convertible c~x$, triggers the new case above when comparing the two argument stacks for $c$, causing $\trueb$ to be returned without evaluating $t$ at all.

\section{An explicitly-scheduled abstract machine} \label{s:scheduling}

\begin{figure}[p]
\begin{flushleft}
{\normalsize
Initial state when testing convertibility of $t_1$ and $t_2$: \hfill
($\alpha \mapsto \{ \mathtt{*} \}$ means that $\alpha$ is always needed)
}
\begin{align*}
&\machstate{\parp{\parp{
\sendp{\alpha}{\convf{\beta}{\gamma}{\nil}}
}{
\sendp{\beta}{\evalf{t_1}{\nil}}
}}{\parp{
\sendp{\gamma}{\evalf{t_2}{\nil}}
}{
KP
}}}{(\alpha \mapsto \{ \mathtt{*} \}), \alpha \cons \nil}
\end{align*}
{\normalsize
Transitions for convertibility processes:
}
\begin{align*}
\machstate{\parp{\sendp{\alpha}{\convf{\beta}{\beta}{\xi}}}{P}}
          {W, \alpha \cons Q} & \red
\machstate{\parp{\sendp{\alpha}{\trueb}}{P}}
          {\finish(\alpha, W, Q)}
\\
\machstate{\parp{\sendp{\alpha}{\convf{\beta}{\beta'}{\xi}}}{P}}
          {W, \alpha \cons Q} & \red
\machstate{\parp{\sendp{\alpha}{\convvf{\recvp{\beta}}{\recvp{\beta'}}{\xi}}}{P}}
          {W, Q \cons \alpha}
\quad \text{if $\beta \not= \beta'$}
\\
\intertext{\normalsize Obtaining the values to be compared:}
\machstate{\parpp{\sendp{\alpha}{\convvf{\recvp{\beta}}{E'}{\xi}}}
                 {\sendp{\beta}{v}}{P}}
          {W, \alpha \cons Q} &\red
\machstate{\parpp{\sendp{\alpha}{\convvf{v}{E'}{\xi}}}
                 {\sendp{\beta}{v}}{P}}
          {W, Q \cons \alpha}
\\
\machstate{\parpp{\sendp{\alpha}{\convvf{E}{\recvp{\beta'}}{\xi}}}
                 {\sendp{\beta'}{v'}}{P}}
          {W, \alpha \cons Q} &\red
\machstate{\parpp{\sendp{\alpha}{\convvf{E}{v'}{\xi}}}
                 {\sendp{\beta'}{v'}}{P}}
          {W, Q \cons \alpha}
\\
\machstate{\parp{\sendp{\alpha}{\convvf{\recvp{\beta}}{\recvp{\beta'}}{\xi}}}{P}}
          {W, \alpha \cons Q} &\red
\machstatetwolines
{\parp{\sendp{\alpha}{\convvf{\recvp{\beta}}{\recvp{\beta'}}{\xi}}}{P}}
{\wakeup(\alpha, \beta, \wakeup(\alpha, \beta', W, Q))}
\\
\machstate{\parp{\sendp{\alpha}{\convvf{\recvp{\beta}}{v'}{\xi}}}{P}}
          {W, \alpha \cons Q} &\red
\machstate{\parp{\sendp{\alpha}{\convvf{\recvp{\beta}}{v'}{\xi}}}{P}}
          {\wakeup(\alpha, \beta, W, Q)}
\\
\machstate{\parp{\sendp{\alpha}{\convvf{v}{\recvp{\beta'}}{\xi}}}{P}}
          {W, \alpha \cons Q} &\red
\machstate{\parp{\sendp{\alpha}{\convvf{v}{\recvp{\beta'}}{\xi}}}{P}}
          {\wakeup(\alpha, \beta', W, Q)}
\\
\intertext{\normalsize Comparing two values:}
\machstate{\parp{\sendp{\alpha}{\convvf{\deepclos{x}{t}{e}{y}{\delta}}
                                       {\deepclos{x'}{t'}{e'}{y'}{\delta'}}
                                       {\xi}}}{P}}
          {W, \alpha \cons Q} & \red
\machstate{\parp{\sendp{\alpha}{\convf{\delta}{\delta'}{((y, y') \cons \xi)}}}{P}}
          {W, Q \cons \alpha}
\\
\machstate{\parp{\sendp{\alpha}{\convvf{\neutralv{x}{s}}
                                       {\neutralv{x'}{s'}}
                                       {\xi}}}{P}}
          {W, \alpha \cons Q} &\red
\machstate{\parp{\sendp{\alpha}{\cstkf{s}{s'}{\xi}}}{P}}
          {W, Q \cons \alpha}
\\& \notag \qquad \text{if $(x, x') \in \xi$ and $\listlength{s} = \listlength{s'}$}
\\
\machstate{\parp{\sendp{\alpha}{\convvf{\constv{c}{s}{\delta}}{\constv{c'}{s'}{\delta}}{\xi}}}{P}}
          {W, \alpha \cons Q}
&\red
\machstate{\parp{\sendp{\alpha}{\trueb}}{P}}
          {\finish(\alpha, W, Q)}
\\
\machstate{\parp{\sendp{\alpha}{\convvf{\constv{c}{s}{\delta}}{\constv{c'}{s'}{\delta'}}{\xi}}}{P}}
          {W, \alpha \cons Q}
&\red
\freshp{\beta\gamma}{
\begin{array}[t]{@{~}l}
    \sendp{\alpha}{\recvp{\beta} \choice \recvp{\gamma}}
\\ \parallel
    \sendp{\beta}{\convvf{\constv{c}{s}{\delta}}{\recvp{\delta'}}{\xi}}
\\ \parallel
    \sendp{\gamma}{\convvf{\recvp{\delta}}{\constv{c'}{s'}{\delta'}}{\xi}}
   \parallel P
\\ \schedstate{W[\beta \upd \{ \alpha \}, \gamma \upd \{ \alpha \}], Q \cons \beta \cons \gamma}
\end{array}
}
\\&\qquad\text{if $c \neq c'$ or $\listlength{s} \not= \listlength{s'}$}
\\
\machstate{\parp{\sendp{\alpha}{\convvf{\constv{c}{s}{\delta}}{\constv{c}{s'}{\delta'}}{\xi}}}{P}}
          {W, \alpha \cons Q}
&\red
\begin{array}[t]{@{~}l}
\freshp{\beta\gamma\eta\zeta}{}
\\
    \sendp{\alpha}{\recvp{\eta} \biasedor \recvp{\zeta}} \parallel \sendp{\zeta}{\recvp{\beta} \choice \recvp{\gamma}}
\\ \parallel
    \sendp{\beta}{\convvf{\constv{c}{s}{\delta}}{\recvp{\delta'}}{\xi}}
\\ \parallel
    \sendp{\gamma}{\convvf{\recvp{\delta}}{\constv{c'}{s'}{\delta'}}{\xi}}
\\
   \parallel
    \sendp{\eta}{\cstkf{s}{s'}{\xi}} 
   \parallel P
\\ \schedstatetwolines
     {W[\eta \upd \{ \alpha \}, \zeta \upd \{ \alpha \}, \beta \upd \{ \zeta \}, \gamma \upd \{ \zeta \}]}
     {Q \cons \eta \cons \beta \cons \gamma}
\end{array}
\\&\qquad\text{if $\listlength{s} = \listlength{s'}$}
\\
\machstate{\parp{\sendp{\alpha}{\convvf{\constv{c}{s}{\delta}}{v'}{\xi}}}{P}}
          {W, \alpha \cons Q}
&\red
\machstate{\parp{\sendp{\alpha}{\convvf{\recvp{\delta}}{v'}{\xi}}}{P}}
          {W, Q \cons \alpha}
\\
\machstate{\parp{\sendp{\alpha}{\convvf{v}{\constv{c'}{s'}{\delta'}}{\xi}}}{P}}
          {W, \alpha \cons Q}
&\red
\machstate{\parp{\sendp{\alpha}{\convvf{v}{\recvp{\delta'}}{\xi}}}{P}}
          {W, Q \cons \alpha}
\\
\machstate{\parp{\sendp{\alpha}{\convvf{v_1}{v_2}{\xi}}}{P}}
          {W, \alpha \cons Q} &\red
\machstate{\parp{\sendp{\alpha}{\falseb}}{P}}
          {\finish(\alpha, W, Q)}
\\&\qquad \text{in all other cases}
\end{align*}
\end{flushleft}
\caption{The explicitly-scheduled abstract machine, part 1.}
\label{f:scheduled1}
\end{figure}

\begin{figure}[p]
\begin{flushleft}
{\normalsize
Comparing two argument stacks:
}
\begin{align*}
\machstate{\parp{\sendp{\alpha}{\cstkf{\nil}{\nil}{\xi}}}{P}}
          {W, \alpha \cons Q} &\red
\machstate{\parp{\sendp{\alpha}{\trueb}}{P}}
          {\finish(\alpha, W, Q)}
\\
\machstate{\parp{\sendp{\alpha}{\cstkf{(\beta \cons s)}{(\beta' \cons s')}{\xi}}}{P}}
          {w, \alpha \cons Q} &\red
\freshp{\gamma\eta}{
\begin{array}[t]{@{~}l}
  \sendp{\alpha}{\recvp{\gamma} \wedge \recvp{\eta}}
\\ \parallel
  \sendp{\gamma}{\convf{\beta}{\beta'}{\xi}}
\\ \parallel
  \sendp{\eta}{\cstkf{s}{s'}{\xi}}
 \parallel P
\\
  \schedstate{W[\gamma \upd \{\alpha\}, \eta \upd \{\alpha\}], Q \cons \gamma \cons \eta}
\end{array}}
\\
\intertext{\normalsize Transitions for evaluation processes:}
\machstate{\parp{\sendp{\alpha}{\evalf{t}{e}}}{P}}
          {W, \alpha \cons Q} & \red
\machstate{\parp{\sendp{\alpha}{\reducef{t}{e}{\nil}}}{P}}
          {W, Q \cons \alpha}
\\
\machstate{\parp{\sendp{\alpha}{\reducef{(t\ u)}{e}{s}}}{P}}
          {W, \alpha \cons Q} &\red
\freshp{\beta}{\machstate{\parpp{\sendp{\alpha}{\reducef{t}{e}{(\beta \cons s)}}}
                                {\sendp{\beta}{\evalf{u}{e}}}
                                {P}}
                         {W, Q \cons \alpha}}
\\
\machstate{\parp{\sendp{\alpha}{\reducef{(\lambda x. t)}{e}{s}}}{P}}
          {W, \alpha \cons Q} &\red 
\freshp{\gamma \delta}{\begin{array}[t]{@{~}l}
     {\sendp{\alpha}{\applyf{\deepclos{x}{t}{e}{y}{\delta}}{s}}}
  \\ \parallel
     {\sendp{\delta}{\evalf{t}{(\extenv{e}{x}{\gamma})}}}
  \\ \parallel
     {\sendp{\gamma}{\neutralvar{y}}}
     \parallel
     {P}
  \\ \schedstate{W, Q \cons \alpha}
\end{array}}
\\& \qquad \text{where $y$ is a fresh variable}
\\
\machstate{\parp{\sendp{\alpha}{\reducef{x}{e}{s}}}{P}}
          {W, \alpha \cons Q} &\red
\machstate{\parp{\sendp{\alpha}{\applyf{\recvp{\getenv{e}{x}}}{s}}}{P}}
          {W, Q \cons \alpha}
\quad\text{if $x \in \Dom(e)$}
\\
\machstate{\parp{\sendp{\alpha}{\reducef{x}{e}{s}}}{P}}
          {W, \alpha \cons Q} &\red
\machstate{\parp{\sendp{\alpha}{\applyf{\neutralvar{x}}{s}}}{P}}
          {W, Q \cons \alpha}
\quad\text{if $x \notin \Dom(e)$}
\\
\machstate{\parp{\sendp{\alpha}{\reducef{c}{e}{s}}}{P}}
          {W, \alpha \cons Q} &\red
\machstate{\parp{\sendp{\alpha}{\applyf{(\constvinit{c}{K(c)})}{s}}}{P}}
          {W, Q \cons \alpha}
\\
\intertext{\normalsize Obtaining the  value to be applied:}
\machstate{\parpp{\sendp{\alpha}{\applyf{\recvp{\beta}}{s}}}{\sendp{\beta}{v}}{P}}
          {W, \alpha \cons Q} &\red
\machstate{\parpp{\sendp{\alpha}{\applyf{v}{s}}}{\sendp{\beta}{v}}{P}}
          {W, Q \cons \alpha}
\\
\machstate{\parp{\sendp{\alpha}{\applyf{\recvp{\beta}}{s}}}{P}}
          {W, \alpha \cons Q} &\red
\machstate{\parp{\sendp{\alpha}{\applyf{\recvp{\beta}}{s}}}{P}}
          {\wakeup(\alpha, \beta, W, Q)}
\\
\intertext{\normalsize Applying a value to a stack:}
\machstate{\parp{\sendp{\alpha}{\applyf{v}{\nil}}}{P}}
          {W, \alpha \cons Q} &\red
\machstate{\parp{\sendp{\alpha}{v}}{P}}{ \finish(\alpha, W, Q)}
\\
\machstate{\parp{\sendp{\alpha}{\applyf{\deepclos{x}{t}{e}{y}{\delta}}{(\beta \cons s)}}}{P}}
          {W, \alpha \cons Q} &\red
\machstate{\parp{\sendp{\alpha}{\reducef{t}{(\extenv{e}{x}{\beta})}{s}}}{P}}
          {W, Q \cons \alpha}
\\
\machstate{\parp{\sendp{\alpha}{\applyf{\neutralv{x}{s'}}{s}}}{P}}
          {W, \alpha \cons Q} &\red
\machstate{\parp{\sendp{\alpha}{\neutralv{x}{(s' \append s)}}}{P}}
          {  \finish(\alpha, W, Q) }
\\
\machstate{\parp{\sendp{\alpha}{\applyf{(\constv{c}{s'}{\delta})}{s}}}{P}}
          {W, \alpha \cons Q} &\red
\freshp{\gamma}{
     \machstate{\parpp{\sendp{\alpha}{\constv{c}{(s' \append s)}{\gamma}}}
                      {\sendp{\gamma}{\applyf{\recvp{\delta}}{s}}}
                      {P}}
               {\finish(\alpha, W, Q)}}
\\
\intertext{\normalsize Transitions for Boolean connectors (symmetrical rules for $\wedge$ and $\choice$ omitted):}
\machstate{\parpp{\sendp{\alpha}{\recvp{\beta} \choice \recvp{\gamma}}}
                 {\sendp{\beta}{v}}
                 {P}}
          {W, \alpha \cdot Q} &\red
\machstate{\parpp{\sendp{\alpha}{v}}{\sendp{\beta}{v}}{P}}
          {\finish(\alpha, \unneed(\alpha, \gamma, W, Q))}
\\
\machstate{\parpp{\sendp{\alpha}{\recvp{\beta}}}{\sendp{\beta}{v}}{P}}
          {W, \alpha \cons Q} &\red
\machstate{\parpp{\sendp{\alpha}{v}}{\sendp{\beta}{v}}{P}}
          { \finish(\alpha, W, Q) }
\\
\machstate{\parpp{\sendp{\alpha}{\recvp{\beta} \wedge \recvp{\gamma}}}{\sendp{\beta}{\falseb}}{P}}
          {W, \alpha \cons Q} &\red
\machstate{\parpp{\sendp{\alpha}{\falseb}}{\sendp{\beta}{\falseb}}{P}}
          {\finish(\alpha, \unneed(\alpha, \gamma, W, Q))}
\\
\machstate{\parpp{\sendp{\alpha}{\recvp{\beta} \wedge \recvp{\gamma}}}{\sendp{\beta}{\trueb}}{P}}
          {W, \alpha \cons Q} &\red
\machstate{\parpp{\sendp{\alpha}{\recvp{\gamma}}}{\sendp{\beta}{\trueb}}{P}}
          {\wakeup(\alpha, \gamma, W, Q)}
\\
\machstate{\parpp{\sendp{\alpha}{\recvp{\beta} \biasedor \recvp{\gamma}}}{\sendp{\beta}{\trueb}}{P}}
          {W, \alpha \cons Q} &\red
\machstate{\parpp{\sendp{\alpha}{\trueb}}{\sendp{\beta}{\trueb}}{P}}
          {\finish(\alpha, \unneed(\alpha, \gamma, W, Q))}
\\
\machstate{\parpp{\sendp{\alpha}{\recvp{\beta} \biasedor \recvp{\gamma}}}{\sendp{\beta}{\falseb}}{P}}
          {W, \alpha \cons Q} &\red
\machstate{\parpp{\sendp{\alpha}{\recvp{\gamma}}}{\sendp{\beta}{\falseb}}{P}}
          {\wakeup(\alpha, \gamma, W, Q)}
\\
\machstate{\parpp{\sendp{\alpha}{\recvp{\beta} \biasedor \recvp{\gamma}}}{\sendp{\gamma}{v}}{P}}
          {W, \alpha \cdot Q} &\red
\machstate{\parpp{\sendp{\alpha}{v}}{\sendp{\gamma}{v}}{P}}
          {\finish(\alpha, \unneed(\alpha, \beta, W, Q))}
\end{align*}
\end{flushleft}

\caption{The explicitly-scheduled abstract machine, part 2.}
\label{f:scheduled2}

\end{figure}

\afterpage{\clearpage}

The evaluation and convertibility testing functions of~\secref{s:reduction} and~\secref{s:convertibility} do not enforce laziness: they allow evaluations to start reducing as soon as the corresponding processes are created, before we know that their results are needed.  We now make scheduling explicit in these functions, specifying which processes should be reduced at each step, so that processes are not executed before their results are needed, and the executions of active processes are interleaved fairly.

The explicitly-scheduled rules are shown in figures~\ref{f:scheduled1} and~\ref{f:scheduled2}.  They can be viewed as the transitions of an abstract machine whose states are triples $\machstate{P}{W,Q}$ or, equivalently, as reduction rules for processes $P$ annotated with scheduling information $W, Q$.  In both cases,
\begin{itemize}
\item $P$ is the parallel composition of a set of elementary processes $\sendp{\alpha_1}{E_1} \parallel \cdots \parallel \sendp{\alpha_n}{E_n}$, identified by their channels $\alpha_i$.
\item $W$ is a map from channels to sets of channels.  It records which processes are waiting on the result of other processes: $W(\alpha) = \{\beta_1, \ldots, \beta_n\}$ means that the processes $\beta_1, \ldots, \beta_n$ are blocked waiting for the process $\alpha$ to send a value.
\item $Q$ is a queue of channels representing the active processes: the processes whose results are needed to advance the resolution of the current convertibility problem.  Inactive processes are never scheduled for execution.  However, a process can alternate between the ``inactive'' and ``active'' states as the convertibility problem progresses. There are no duplicates in this queue, ensuring fairness between the active processes.
\end{itemize}

Many of the abstract machine transitions simply perform round-robin execution of the active processes.  They have the following shape:
$$
\machstate{\sendp{\alpha}{E} \parallel P}{W, \alpha \cons Q}
\red
\machstate{P' \parallel P}{W, Q \cons \alpha}
$$
where $\sendp{\alpha}{E} \red P'$ is one of the reduction steps for the evaluation or convertibility functions of~\secref{s:reduction} and~\secref{s:convertibility}.  The reduction step is performed because the process $\alpha$ is active and in head position in the queue $Q$.  The process is then moved to the end of $Q$.  The other processes $P$ and the wait map $W$ are unchanged.  The net effect of these transitions is to step through the executions of the active processes in round-robin manner.

When a new process is created, it can be either in the inactive state or in the active state.  For example, in the rule that reduces a function application
$$
\machstate{\parp{\sendp{\alpha}{\reducef{(t\ u)}{e}{s}}}{P}}
          {W, \alpha \cons Q} \red
\freshp{\beta}{\machstate{\parpp{\sendp{\alpha}{\reducef{t}{e}{(\beta \cons s)}}}
                                {\sendp{\beta}{\evalf{u}{e}}}
                                {P}}
                         {W, Q \cons \alpha}}
$$
the process $\sendp{\beta}{\evalf{u}{e}}$ that computes the value of the function argument $u$ is initially inactive, since we do not need its value right away.  It will become active when the value $\recvp{\beta}$ is needed.  In contrast, the rule that tests the convertibility of two nonempty stacks,
$$
\machstate{\parp{\sendp{\alpha}{\cstkf{(\beta \cons s)}{(\beta' \cons s')}{\xi}}}{P}}
          {Q, \alpha \cons W} \red
\freshp{\gamma\eta}{
\begin{array}[t]{@{~}l}
  \sendp{\alpha}{\recvp{\gamma} \wedge \recvp{\eta}}
\\ \parallel
  \sendp{\gamma}{\convf{\beta}{\beta'}{\xi}}
\\ \parallel
  \sendp{\eta}{\cstkf{s}{s'}{\xi}}
 \parallel P
\\
  \schedstate{W[\gamma \upd \{\alpha\}, \eta \upd \{\alpha\}], Q \cons \gamma \cons \eta}
\end{array}}
$$
creates two new convertibility processes, $\gamma$ and $\eta$, which need to start executing right away, while $\alpha$ need to wait for them to produce Boolean values.  Therefore, $\gamma$ and $\eta$ are added to $Q$, and $\alpha$ is recorded (in $W$) as waiting both on $\gamma$ and on $\eta$.

When an active process reduces to a value, all the processes waiting for this value must be restarted.  This is performed by the $\finish$ operation:
$$
\finish(\alpha, W, Q) =
  (W[\alpha \upd \emptyset], Q \append (W(\alpha) \setminus Q))
$$
A typical use is the application of a value to an empty stack:
$$
\machstate{\parp{\sendp{\alpha}{\applyf{v}{\nil}}}{P}}
          {W, \alpha \cons Q} \red
\machstate{\parp{\sendp{\alpha}{v}}{P}}{ \finish(\alpha, W, Q)}
$$
The effect of $\finish$ is to restart all the processes that were blocked while reading from $\alpha$, adding them to the queue of active processes.  Then, $W(\alpha)$ is set to $\emptyset$ since no process remains waiting on $\alpha$.  The process $\alpha$ is not added back to $Q$, since it has terminated and does not need to be scheduled ever again.

Symmetrically, the $\wakeup(\alpha, \beta, W, Q)$ operation suspends the process $\alpha$ until the process $\beta$ has produced a value.
\begin{align*}
\wakeup(\alpha, \beta, W, Q) &=
   (\updatew{W}{\beta}{W(\beta) \cup \{ \alpha \}}, Q)
   &&\text{if $W(\beta) \neq \emptyset$}
\\
\wakeup(\alpha, \beta, W, Q) &=
   (\updatew{W}{\beta}{\{ \alpha \}}, Q \cons \beta)
   &&\text{if $W(\beta) = \emptyset$}
\end{align*}
A typical use of $\wakeup$ is the rule for applying a value read from a channel:
\begin{equation*}
\machstate{\parp{\sendp{\alpha}{\applyf{\recvp{\beta}}{s}}}{P}}
          {W, \alpha \cons Q} \red
\machstate{\parp{\sendp{\alpha}{\applyf{\recvp{\beta}}{s}}}{P}}
          {\wakeup(\alpha, \beta, W, Q)}
\tag{i}
\end{equation*}
The process $\sendp{\alpha}{\applyf{\recvp{\beta}}{s}}$ needs to receive a value from channel $\beta$ before it can proceed.  Therefore, $\alpha$ becomes inactive, $\beta$ becomes active if it was not already, and a dependency of $\alpha$ on $\beta$ is added to the map $W$.

Note that the $\wakeup$ operation must not be performed if the desired value is already available.  To ensure this, for each rule like (i) above, we have a companion rule:
\begin{equation*}
\machstate{\parpp{\sendp{\alpha}{\applyf{\recvp{\beta}}{s}}}{\sendp{\beta}{v}}{P}}
          {W, \alpha \cons Q} \red
\machstate{\parpp{\sendp{\alpha}{\applyf{v}{s}}}{\sendp{\beta}{v}}{P}}
          {W, Q \cons \alpha}
\tag{ii})
\end{equation*}
where the value $v$ produced by process $\beta$ is directly transferred to process $\alpha$.  (The convention we follow for the abstract machine is that transitions are determined by the first rule that matches.  Since rule (ii) appears before rule (i) in figure~\ref{f:scheduled2}, (ii) takes precedence over (i).)

A process may need to wait for several processes to produce their values.  This is the case in the rule that obtains two values to be compared:
$$
\machstate{\parp{\sendp{\alpha}{\convvf{\recvp{\beta}}{\recvp{\beta'}}{\xi}}}{P}}
          {W, \alpha \cons Q}
\red
\machstate{\parp{\sendp{\alpha}{\convvf{\recvp{\beta}}{\recvp{\beta'}}{\xi}}}{P}}
          {\wakeup(\alpha, \beta, \wakeup(\alpha, \beta', W, Q))}
$$
Here, both $\beta$ and $\beta'$ need to be restarted, and $\alpha$ must be marked as waiting on both $\beta$ and $\beta'$.

Finally, when evaluating Boolean connectors, the value of an active process may become unneeded.  For example, if we are evaluating $\sendp{\alpha}{\recvp{\beta} \land \recvp{\gamma}}$ and the process $\beta$ produces $\falseb$, we no longer need the value of $\gamma$.  Therefore, we update the scheduling state to reflect this fact before producing $\falseb$ on $\alpha$:
$$
\machstate{\parpp{\sendp{\alpha}{\recvp{\beta} \wedge \recvp{\gamma}}}
                 {\sendp{\beta}{\falseb}}
                 {P}}
          {W, \alpha \cdot Q} \red
\machstate{\parpp{\sendp{\alpha}{\falseb}}{\sendp{\beta}{\falseb}}{P}}
          {\finish(\alpha, \unneed(\alpha, \gamma, W, Q))}
$$
The $\unneed$ operation is defined recursively as
\begin{align*}
\unneed(\alpha, \beta, W, Q) &=
   (\updatew{W}{\beta}{W(\beta) \setminus \{ \alpha \}}, Q)
   \qquad \text{if $W(\beta) \neq \{ \alpha \}$}
\\
\unneed(\alpha, \beta, W, Q) &=
  \unneed(\beta, \gamma_1, \dots, \unneed(\beta, \gamma_n, \updatew{W}{\beta}{\emptyset}, Q \setminus \{ \beta \}))
\\
& \qquad \text{if $W(\beta) = \{ \alpha \}$ and $\{ \gamma \mid \beta \in W(\gamma) \} = \{ \gamma_1, \dots, \gamma_n \}$}
\end{align*}
The dependency of $\alpha$ on $\beta$ is removed from $W$.  Moreover, if $\alpha$ was the only process waiting on $\beta$ to produce a value, the execution of $\beta$ is stopped by removing $\beta$ from the queue $Q$ of active processes, and we recursively call $\unneed(\beta, \gamma)$ on all processes $\gamma$ that $\beta$ was waiting for.  This is similar to removing a reference in a reference counting system.  The $\unneed$ operation can be implemented efficiently by using a doubly linked list for $Q$ and storing $W$ as a pair of maps in both directions.

\section{Rocq proof} \label{s:rocq-proof}

We formally verified the partial correctness of our concurrent convertibility test, using the Rocq interactive theorem prover. The formalization includes the main algorithm presented in \secref{s:convertibility}, without the extensions shown in \secref{s:extensions}. It also incorporates extensions to the core $\lambda$-calculus to handle data constructors and pattern-matching. The proof only proves partial correctness; that is, if the algorithm produces a result, then the result is correct.  It does not prove termination.  The proof does not formalize scheduling either, because it is unnecessary for proving correctness.

The reduction part of the algorithm is formulated in the style of \secref{s:reduction}, as transitions over sets of concurrent threads, leaving scheduling unspecified.  Since the sharing of convertibility processes shown in \secref{s:sharing-convertibility-processes} is not included, all convertibility processes of \secref{s:convertibility} are replaced by a tree of Boolean operations.  The leaves of this tree are convertibility tests between two channels.

The development is included as an artifact for this paper.  It sums up to about 10000 lines of Rocq in total, which makes it a moderately-sized proof. Here is the final theorem:
\begin{lstlisting}[language=coq]
Lemma all_correct :
  forall defs t1 t2 st b,
    defs_wf defs ->
    closed_at t1 0 -> closed_at t2 0 ->
    dvar_below (length defs) t1 -> dvar_below (length defs) t2 ->
    star step (init_conv defs t1 t2) (cthread_done b, st) ->
    reflect (convertible (betaiota defs) t1 t2) b.
\end{lstlisting}
It expresses that, given well-formed constant definitions "defs" and two closed terms "t1" and "t2" that reference only constants defined in "defs", if we start the convertibility abstract machine in the initial state corresponding to "defs", "t1" and "t2", and if it stops after a finite number of transitions on a final "cthread\_done" state carrying the Boolean result "b", then "b" is "true" if and only if "t1" and "t2" are convertible.  The "betaiota" relation, which should actually be called "betadelta", is the union of beta reduction and unrolling of defined constants.
Note that this does not guarantee termination nor the absence of errors or deadlocks: we have not proved that a "cthread\_done" state will be reached. However, this guarantees that once this state is reached, then the result thus obtained is correct.

Unsurprisingly, one of the more delicate aspects of the Rocq development is the representation of variables in terms.  We use de Bruijn indices for the inputs of the convertibility test (such as the terms "t1" and "t2" above) and named variables for evaluated terms.  De Bruijn indices are easier to work with, but named variables allow for sharing without the need for explicit weakenings. This approach necessitates generating fresh variable names in the state of the reduction, and proving numerous invariants justifying that the generated variables are indeed fresh.

One limitation of the Rocq proof is that, for a reduction step from a configuration representing a term $t$ to a configuration representing a term $t'$, it only shows that $t$ and $t'$ are convertible ($t \convertible t'$) but not that $t$ reduces to $t'$ ($t \redplus t'$).  Knowing that $t \convertible t'$ is enough to prove the partial correctness of the convertibility checking algorithm. However, we would need to know that $t \redplus t'$ in order to prove that convertibility always terminates when given two strongly normalizing terms.  We previously had a proof of $t \redplus t'$ for an earlier, simpler version of our call-by-need evaluator, but the proof was so complex and difficult to extend that we switched to the simpler proof of $t \convertible t'$.

\section{Performance analysis} \label{s:performance-analysis}

It is not obvious how to characterize the performance of an algorithm for convertibility checking.  There is no useful upper bound as a function of the size $n$ of the input terms: the convertibility problem is TOWER-complete even when restricted to simply-typed terms \cite{DBLP:journals/tcs/Statman79a} \cite{lmcs:11344}.  \citet{PHD:Condoluci} gives an $\bigO(mn)$ complexity bound, where $m$ is the number of reductions needed to fully normalize the input terms and $n$ is their size.  Since our algorithm avoids computing normal forms as much as it can, we would prefer a bound that does not involve $m$.

\begin{figure}
\begin{mathpar}
\inferrule*[right=red-l]
  {t \redbd t' \\ t' \convertible u : B}
  {t \convertible u : B}
\and
\inferrule*[right=red-r]
  {u \redbd u' \\ t \convertible u' : B}
  {t \convertible u : B}
\\
\inferrule*[right=lam]
  {t \convertible u[y \becomes x] : B}
  {\lam x. t \convertible \lam y. u : B}
\and
\inferrule*[right=lam-var]
  { }
  {\lam x. t \convertible y~u_1~\cdots~u_n : \falseb}
\and
\inferrule*[right=var-lam]
  { }
  {x~t_1~\cdots~t_n \convertible \lam y. u : \falseb}
\\
\inferrule*[right=var-1]
  {t_1 \convertible u_1 : \trueb \quad\ldots\quad t_n \convertible u_n : \trueb}
  {x~t_1~\cdots~t_n \convertible x~u_1~\cdots~u_n : \trueb}
\and
\inferrule*[right=var-2]
  {x \not= y}
  {x~t_1~\cdots~t_n \convertible y~u_1~\cdots~u_m : \falseb}
\and
\inferrule*[right=var-3]
  {t_i \convertible u_i : \falseb}
  {x~t_1~\cdots~t_n \convertible x~u_1~\cdots~u_n : \falseb}
\and
\inferrule*[right=const]
  {t_1 \convertible u_1 : \trueb \quad\ldots\quad t_n \convertible u_n : \trueb}
  {c~t_1~\cdots~t_n \convertible c~u_1~\cdots~u_n : \trueb}
\end{mathpar}

\caption{The inference rules for the convertibility judgment $t \convertible u : B$}
\label{f:rules}

\end{figure}

To this end, we take a step back from the details of the algorithm and view convertibility checking as a \emph{proof search} problem.  Given two terms $t$ and $u$, we aim to derive the judgment $t \convertible u : B$ where the Boolean $B$ is $\trueb$ if $t$ and $u$ are convertible and $\falseb$ otherwise.  The inference rules for this judgment are given in figure~\ref{f:rules}. Rules \RightTirName{red-l} and \RightTirName{red-r} correspond to performing one reduction step $\redbd$ in $t$ or in $u$, either beta-reduction ($\beta$) or unrolling of a defined constant ($\delta$).  The other rules follow the structure of the two terms $t$ and $u$.  Rule \RightTirName{const} is the ``shortcut'' for proving that two applications of a defined constant $c$ are convertible without unrolling $c$.

A goal $t \convertible u : B$ generally admits multiple proofs, but some proofs are smaller than others.  For example, $c~x \convertible c~x : \trueb$ has a proof of size 2 (rules \RightTirName{const} and \RightTirName{var-1}) and other, longer proofs obtained by first unrolling $c$ on both sides (rules \RightTirName{red-l} and \RightTirName{red-r}).

The algorithm of \secref{s:scheduling} can be viewed as a \emph{breadth-first search} of the tree of possible proofs.  For example, given the problem $c~t_1~\cdots~t_n \convertible c~u_1~\cdots~u_n : B$, the algorithm searches in parallel for three kinds of possible proofs, those ending with rule~\RightTirName{const}, those ending with rule~\RightTirName{red-l}, and those ending with rule~\RightTirName{red-r}. In contrast, other convertibility checkers, such as Rocq's, perform a \emph{depth-first search} for a proof of convertibility, guided by heuristics.

The first author proved that
if there exists a proof of $t \convertible u : B$ of size $s$, the algorithm of \secref{s:scheduling} terminates in time $\bigO((k + 1)^{2s})$, where $k \ge 1$ is the maximal arity of variable and constant applications in the original terms $t, u$
\cite[chapter 12]{courant:tel-04884688}.
In other words, our algorithm is exponential in the size $s$ of the smallest proof of (non-)convertibility.  In contrast, convertibility checkers based on depth-first search can take time unbounded by any function of $s$, since they can perform an arbitrarily large amount of computation before finding a proof.  %

The complexity argument above needs to be made more precise: the size $s$ of the smallest convertibility proof depends crucially on the inference rules and the reduction strategy used.  While the inference rules in figure~\ref{f:rules} are somewhat canonical, the strategy used by $\redbd$ reductions in rules \RightTirName{red-l} and \RightTirName{red-r} has a huge impact on the size $s$ of convertibility proofs.  For instance, using weak call-by-name or weak call-by-value can result in convertibility proofs that are exponentially bigger (or worse) than those obtained using weak call-by-need; and using optimal reduction \cite{DBLP:conf/popl/Lamping90} could lead to even smaller proofs.  

To clarify this dependency on the reduction strategy used, the complexity argument
of \citet[chapter 12]{courant:tel-04884688}
is formulated in terms of an \emph{effective reduction structure}, which is an abstract presentation of graph reduction.  Therefore, the complexity argument is independent of the details of our call-by-need evaluator, and only relies on the sharing properties of graph reduction.

The size $s$ of a convertibility proof can be decomposed as $s = r + f + b$, where $r$ is the number of reduction steps, $f$ the number of ``forced'' convertibility steps (those where only one rule applies to the current goal), and $b$ the number of ``branching'' convertibility steps (those where several rules apply to the current goal).  Obviously, the branching convertibility steps are those responsible for the exponential overhead of our algorithm.  We conjecture that there exist scheduling strategies for which our algorithm has a complexity bound of $\bigO((r + f) K^{b})$ for some constant $K \ge 2$, instead of $\bigO(K^{r + f + b})$ as in the analysis above.  The idea is to use non-uniform scheduling of processes, where each process has a a share of CPU time, with all shares summing to 1.  Each process is scheduled with a frequency proportional to its share.  A convertibility process that performs a branching step would divide its share among the processes that it creates.  The effect of this scheduling policy would be to slow down the exponential explosion in the number of processes caused by branching steps, giving more time to reduction and convertibility processes created earlier.

\section{Experimental evaluation}  \label{s:experimental}

For the experimental evaluation, we used two OCaml implementations of our convertibility checker, which corresponds to the version verified in Rocq extended with inductive types and fixed points.  The implementation named ``Full'' in the following follows the algorithm of \secref{s:scheduling} and implements the sharing of convertibility processes described in \secref{s:sharing-convertibility-processes}.  The implementation named ``Simple'' in the following uses an earlier version of our algorithm that does not share convertibility processes and uses a heuristic to determine which side to unfold first when encountering different head constants, instead of trying both unfoldings in parallel.  Variables in the input terms are represented by the type $"string"$, which comes with additional costs compared to Rocq's internal de Bruijn indices. Neither implementation performs the pre-sharing of subterms described in \secref{s:sharing-subterms}.  On the Rocq side, we instrumented Rocq's convertibility checker so that it prints the time taken. (This is much more precise than just relying on Rocq's $"Time"$ command, which also accounts for other aspects of type checking.) We used Rocq 8.15.2, extended with these changes to the convertibility checker. Moreover, both Rocq and our implementation were compiled by OCaml 4.12.1, and the measurements were performed on a Intel Core i7-1165G7 2.80GHz CPU and 2x 16GiB SODIMM DDR4 Synchronous 3200 MHz (0.3 ns) RAM, running Linux 5.15.74 with NixOS 22.05.

The test cases we used are described and commented below, while the time measurements are shown in figure~\ref{f:convcompare}. On all the test cases, the timings are quite small, because larger inputs would cause stack overflows, and only one digit is significant.  However, these rough measurements are already sufficient to spot nonlinear behaviors.

The test cases use the following defined constants:

\begin{lstlisting}[language=coq]
Fixpoint exp2 n := match n with O => 1 | S n => 2 * exp2 n end.
Definition zero (n : nat) := 0.

Inductive tree := L : tree | N : tree -> tree -> tree.
Fixpoint perfect n t := match n with O => t | S n => perfect n (N t t) end.
Fixpoint ldepth t := match t with L => 0 | N t1 t2 => S (ldepth t1) end.
Fixpoint ldepth2 t := match t with L => 0 | N t1 t2 => ldepth2 t + 1 end.

Definition pair1 n := (is_zero n, n).   Definition pair2 n := (n, is_zero n).

Definition f0 (n : nat) := n.           Definition f1 n := f0 (f0 n).
Definition f2 n := f1 (f1 n).           Definition f3 n := f2 (f2 n).
Definition f4 n := f3 (f3 n).
\end{lstlisting}

\begin{figure} \small
  \newcommand{\sci}[2]{\ensuremath{#1 \times 10^{#2}}}
  \newcommand{\ffourthirty}{\underbrace{"f4"(\cdots("f4"~("f4"~0)) \cdots )}_{\text{30 applications of "f4"}}}
  \def\arraystretch{1.2}
  \begin{center}
    \begin{tabular}[t]{@{}c|l|ll|ll@{}}
      Test case & Rocq & \multicolumn{2}{c|}{Simple} & \multicolumn{2}{c}{Full} \\
                & \emph{time} & \emph{time} & \hspace*{-3mm}\emph{speedup} & \emph{time} & \hspace*{-3mm}\emph{speedup} \\\hline
      $"exp2"~15 \convertible "exp2"~(14 + 1)$ & \sci{3}{-5} & \sci{5}{-5} & -0.22 & \sci{6}{-3} & -2.3 \\
      $"zero"~("exp2"~15) \convertible "zero"~("exp2"~16)$ & 0.14 & \sci{5}{-6} & +4.4 & \sci{2}{-5} & +3.8  \\
      $"ldepth"~("perfect"~15~"L") \convertible "ldepth2"~("perfect"~15~"L")$ & \sci{9}{-5} & \sci{2}{-4} & -0.35 & \sci{5}{-5} & 0.26 \\
      $"perfect"~15~"L" \convertible "perfect"~14~("N"~"L"~"L")$ & 0.018 & 0.013 & +0.14 & \sci{9}{-5} & +2.3 \\
      $("exp2"~15, "false") \not\convertible ("exp2"~16, "true")$ & \sci{4}{-6} & \sci{6}{-6} & -0.18 & \sci{8}{-6} & -0.30 \\
      $("false", "exp2"~15) \not\convertible ("true", "exp2"~16)$ & 0.61 & \sci{1}{-6} & +5.8 & \sci{8}{-6} & +4.9 \\
      $"pair1"~("exp2"~15) \convertible ("false", "exp2"~15)$ & \sci{3}{-5} & \sci{7}{-5} & -0.37 & \sci{2}{-4} & -0.82 \\
      $"pair2"~("exp2"~15) \convertible ("exp2"~15, "false")$ & 0.078 & \sci{5}{-5} & +3.2 & \sci{2}{-4} & +2.6 \\
      $\smash{\ffourthirty} \convertible \smash{\ffourthirty}$
         & \sci{2}{-5} 
         & $\begin{cases} \sci{6}{-5} \\ 0.18 \end{cases}$ \hspace*{-4mm}
         & $\begin{cases} -0.5 \\ -4.0  \end{cases}$ \hspace*{-4mm}
         & 0.15 & -3.9
    \end{tabular}
  \end{center}
  \caption{Timings, in seconds, for the examples of convertibility problems given in the text for Rocq and our two OCaml implementations, Simple and Full.  Speedups are relative to Rocq and are expressed as base-10 logarithms, \emph{i.e.} decimal orders of magnitude.  Higher is faster.  See the main text for the description of the test cases and the explanation of the two results given for the last test.}
  \label{f:convcompare}
\end{figure}

The first two tests, $"exp2"~15 \convertible "exp2"~(14 + 1)$ and $"zero"~("exp2"~15) \convertible "zero"~("exp2"~16)$, focus on the heuristic used when the two head constants are the same. With Rocq, the first test is fast but the second one is slow. In both cases, Rocq attempts to prove the convertibility of the arguments before unfolding the definition of the constant. This is a good strategy for $"exp2"~15 \convertible "exp2"~(14 + 1)$, as it allows Rocq to prove convertibility without expanding $"exp2"$.  However, in the case of $"zero"~("exp2"~15) \convertible "zero"~("exp2"~16)$, Rocq tries and fails to prove the convertibility of $"exp2"~15$ and $"exp2"~16$, which costs a lot of work. Expanding $"zero"$ would have immediately proved convertibility.

Here, our two convertibility checkers get the best of both worlds by doing the work in parallel, and both checks are fast. Our Full checker performs worse on the first test: replacing 15 by $n$ and varying $n$, we experimentally measured $O(n^{2.8})$ complexity instead of the expected $O(n)$.  This is caused by the large amount of unfolding opportunities in the branch where we unfold $"exp2"$, causing an explosion in the number of processes.  This issue could be alleviated by the non-uniform scheduling policy outlined at the end of \secref{s:performance-analysis}.

Next, we will consider terms whose size is exponential in the size of their memory representation, because there is a lot of sharing within the term itself.  The function $"perfect"$ takes an argument $"n"$ and a tree $"t"$ and generates a tree with $2^n$ copies of $"t"$. However, the evaluation only takes time linear in $"n"$ to evaluate, as the subtrees are shared. The definitions $"ldepth"$ and $"ldepth2"$ compute the length of the leftmost branch of their argument, in linear time for $"ldepth"$, and quadratic time for $"ldepth2"$.

Testing the convertibility of $"ldepth"~("perfect"~15~"L")$ and $"ldepth2"~("perfect"~15~"L")$ is fast both in Rocq and with our convertibility checkers because terms are shared; therefore, the computation of both sides takes only quadratic time. However, Rocq is slow to check the convertibility of $"perfect"~15~"L"$ and $"perfect"~14~("N"~"L"~"L")$, because after expanding $"perfect"$, it has to prove the convertibility of the exact same terms multiple times. For the same reason, the Simple version of our convertibility check is slow, but the Full version, which performs more sharing, is fast.

Another interesting test concerns the order in which the arguments of constructors (or identical defined constants) are compared. We consider two very similar tests of non-convertibility, $("exp2"~15, "false") \not\convertible ("exp2"~16, "true")$ and $("false", "exp2"~15) \not\convertible ("true", "exp2"~16)$. With Rocq, the first test is almost instantaneous: Rocq starts by comparing $"true"$ and $"false"$, since Rocq evaluates arguments from right to left. They are different, so the test stops immediately. However, the second test is much slower: Rocq starts by comparing $"exp2"~15$ and $"exp2"~16$, which fails after a long time. With our convertibility checkers, both tests are equally fast: we test the convertibility of the arguments in parallel, so we immediately detect that $"false"$ and $"true"$ are not convertible, and return this result.

Another peculiarity of Rocq is that once a constant is unfolded, it remains unfolded for future tests, preventing us from benefiting from the optimization with folded constants. The next two tests,  $"pair1"~("exp2"~15) \convertible ("false", "exp2"~15)$ and $"pair2"~("exp2"~15) \convertible ("exp2"~15, "false")$ demonstrate the problems this can cause. Again, the two tests are identical except for the order of arguments. In the first test, Rocq first compares $"exp2"~15$ and $"exp2"~15$, which is almost instantaneous thanks to the folded constant optimization. Then, it compares $"is\_zero"~("exp2"~15)$ with $"false"$, which takes only linear time, thanks to Rocq's laziness. However, in the second test, Rocq first compares $"is\_zero"~("exp2"~15)$ with $"false"$, forcing it to unfold $"exp2"$ to prove convertibility. Once this is done, it compares $"exp2"~15$ with a version of $"exp2"~15$ that has already been partially computed and where $"exp2"$ has been unfolded. At this point, it has no way but to expand $"exp2"$ on the other side, and the time taken is exponential. With our convertibility checker, both tests are fast. Indeed, when we unfold a constant, we also keep the original folded value, allowing us to still benefit from the folded constant optimization if we encounter it again.

Of course, this comparison wouldn't be honest if we didn't also show a shortcoming of our own convertibility checker. In the final example, we have an identical term on both sides, but it is deeply nested. Rocq is almost instantaneous there by repeatedly applying the folded constant optimization, but since our checker explores what happens both when unfolding and when not unfolding, it is much slower. The Full version is always slow.  The Simple version uses heuristics to choose which side to unfold when encountering two different head constants, and the speed depends heavily on the unfolding order.  If we choose to always unfold the older constant first, we obtain the result quickly: when we unfold $"f4"$ on one side, then we will repeatedly unfold $"f3"$, $"f2"$, $"f1"$ and then $"f0"$ on that side until that side has only $"f4"$, preventing the folded constant optimisation from applying and spawning new processes until that point. This severely limit the number of total convertibility processes that are created, thus allowing the code to run quite fast, albeit slower than Rocq. On the other hand, if we always unfold the newer constant first (which is often the best choice in Rocq), when we unfold $"f4"$ on one side, we will match this by unfolding $"f4"$ on the other side next, making $"f3"$ appear as the head constant on both sides, making the folded constant optimisation apply again, and so on with $"f2"$, $"f1"$ and $"f0"$, creating in total a very large number of convertibility processes, and thus making the code run very slowly.

However, such examples seem to be quite pathological, and we think they should not happen in practice. Besides, we have a guaranteed complexity of our convertibility test in terms of the shortest existing convertibility proof, which looks like a desirable property that Rocq does not have.

\section{Related work} \label{s:related}

\subsection{Convertibility checking} 

The most advanced algorithms for convertibility testing can be found in the implementations of Agda, Lean, Rocq and other dependently-typed frameworks. However, these algorithms are undocumented and difficult to reconstruct from source code.  The \href{https://github.com/AndrasKovacs/smalltt}{\texttt{smalltt}} project by 
András Kovács is a small, readable implementation of elaboration for dependent types that includes a convertibility checker based on normalization by evaluation.

Among the published work on this topic, the one closest to ours is the MetaRocq (formerly MetaCoq) project, which contains a verified convertibility checker as part of its verified type-checker for the core Rocq language \citep{DBLP:journals/jar/SozeauABCFKMTW20,DBLP:journals/jacm/SozeauFLNTW25}.  Unlike ours, their checker is proved to terminate when given two well-typed terms as inputs, under the assumption that all well-typed terms are strongly normalizing.  
The MetaRocq checker handles defined constants with a fixed strategy (e.g. for $c~t \convertible c~t'$, it always tries $t \convertible t'$ before unfolding $c$) and performs reductions using a Krivine-style machine and a call-by-name strategy, without any support for sharing.

\citet{DBLP:journals/pacmpl/0001OV18} describe another impressive verification of the metatheory of a dependently-typed language, including a constructive proof that convertibility is decidable.  It uses typed reduction and convertibility relations, which facilitate the proof of termination. \citet{DBLP:conf/cpp/AdjedjLMPP24} extend this approach to a verified algorithm for convertibility checking.  The reduction strategy is not specified, and no provision is made for sharing reductions.  Earlier work \citep{DBLP:conf/mpc/AbelCD08} used normalization by evaluation instead of typed reductions, and is therefore restricted to determining $\beta\eta$-convertibility, while \citep{DBLP:journals/pacmpl/0001OV18} handles both $\beta$-convertibility and $\beta\eta$-convertibility.  \citet{DBLP:conf/fscd/Lennon-Bertrand25} compares and relates the typed approach to convertibility checking used in the aforementioned work with the untyped approach that we use in this paper.

The idea that convertibility testing can be performed incrementally by alternating between evaluation to WHNF and comparison of the resulting values goes back at least to \citet{DBLP:journals/scp/Coquand96}. An early implementation of this approach is described by \citet{DBLP:conf/icfp/GregoireL02}.  However, their compiled implementation of WHNF evaluation uses call by value and unrolls constants eagerly, resulting in unnecessary computation.

All the earlier work described above, like our own work, relies on interleaved evaluations and comparisons of values.  \citet{PHD:Condoluci} goes back to the more traditional approach based on normalization of the two terms followed by equality testing of their normal forms, but uses a clever normalization algorithm that exploits sharing in a call-by-value strategy \citep{DBLP:conf/lics/AccattoliCC21} and a clever equality test that takes sharing into account and runs in time linear in the size of the shared representation of the two normal forms \citep{DBLP:conf/ppdp/CondoluciAC19}.

\subsection{Strong call-by-need reduction}

Call-by-need strategies for strong reduction (evaluation under lambda-abstractions) are difficult to define formally and to implement correctly.  \citet{DBLP:journals/pacmpl/BalabonskiBBK17} develop $\lambda_c$, a strong call-by-need calculus that uses explicit substitutions to represent sharing.  To enforce laziness, they need to delay reducing under a lambda-abstraction until all applications of that abstraction have been reduced.  Our enriched function closures $\deepclos{x}{t}{e}{y}{\delta}$ support applying a lambda-abstraction and reducing within its body in any order, which simplifies the presentation.

\citet{DBLP:conf/fscd/BalabonskiLM21} extend $\lambda_c$ with the ability to reduce under lambda-abstractions before applying them if it can be determined that the normal form of the lambda-abstraction will be needed.  They also present an abstract machine that implements these reductions efficiently.  Their approach can perform fewer $\beta$-reductions than ours in some cases.  However, their abstract machine lacks the subterm property and therefore cannot be statically compiled to virtual machine code or native code.

\citet{DBLP:journals/pacmpl/BiernackaCD22} develop a call-by-need normalizer by applying memoization techniques to a call-by-name normalization-by-evaluation function derived from the KN machine of \citet{DBLP:journals/lisp/Cregut07}.  Applying a mechanized CPS transformation to this normalizer, they obtain the RKNL machine, a simple and efficient abstract machine for strong call-by-need evaluation.  While developed independently, our approach to call-by-need normalization described in \secref{s:reduction} is essentially isomorphic to the RKNL machine.

\subsection{Semantics of laziness}

The first formal presentations of call by need and more generally of lazy evaluation used graph reduction; see \citet[part II]{DBLP:books/ph/Jones87} for a survey.  \citet{DBLP:conf/popl/Launchbury93} gave a big-step semantics for lazy evaluation using terms and an explicit store for memoization.  \citet{DBLP:conf/popl/AriolaFMOW95} and \citet{DBLP:journals/jfp/AriolaF97} give small-step semantics using "let" bindings or distinguished $\beta$-redexes to express sharing and laziness of evaluations.  Our process-based notation for lazy / non-strict computations is essentially isomorphic to their "let"-based notation, with parallel processes $\parp{\sendp{\alpha}{t}}{C[\recvp{\alpha}]}$ playing the role of $"let"~x=t~"in"~C[x]$ bindings in \citet{DBLP:conf/popl/AriolaFMOW95}.  We were also inspired by the encoding of the call-by-need weak lambda-calculus in the asynchronous pi-calculus of \citet{DBLP:conf/birthday/Sangiorgi19}, with the difference that Sangiorgi relies on an explicit handshake to delay evaluations until needed, while we rely on an external scheduler.

\section{Conclusions and further work} \label{s:conclusions}

We hope this work sparks renewed interest in convertibility checking, which is a difficult problem that is central to the implementation of type- and proof-checkers.  The lazy, concurrent convertibility checking algorithm described in this paper is novel in several ways: it does not rely on heuristics, it always finds the simplest proof of (non-)convertibility, and its complexity is bounded as a function of the size of the simplest proof.  Admittedly, the bound is exponential, but this is an improvement over heuristics-based sequential algorithms, whose complexity is unbounded in the size of the simplest proof.

This paper focuses on the lambda-calculus with constants.  However, the ideas presented here have been extended to a richer language that includes inductive types with pattern-matching and structural recursion
\cite{courant:tel-04884688}.

As mentioned at the end of \secref{s:performance-analysis}, the worst-case bound of our algorithm, as well as its actual performance on some examples shown in \secref{s:experimental}, could probably be improved by a more sophisticated scheduling of processes that allocates unequal amounts of time to different processes.  Additionally, constant factors could also be reduced by using more clever imperative data structures for scheduling and by compiling evaluation processes to virtual machine code or even to native code.  However, we are skeptical that hardware parallelism can be used to significantly speed up our algorithm, given the slow progress in the area of parallel graph reduction since the 1980s.

The formal proof of \secref{s:rocq-proof} needs more work: to prove that the convertibility checker cannot go wrong or deadlock when given two type-safe terms as input, and that it terminates when given two strongly normalizing terms as input.  The termination proof sounds challenging, especially if we stick to untyped reductions.  Typed reductions in the context of a normalizing type system might provide a simpler proof.  However, this would make the convertibility checker specific to a given type system.

Our convertibility checker can easily be instrumented to generate a trace of the nonobvious unrolling decisions it made.  Using this trace, (non-)convertibility can then be rechecked by a simpler, purely sequential algorithm.  This approach could facilitate the integration of our convertibility checker into an existing proof checker.  Additionally, convertibility traces can  be cached to improve proof checking times when some proof terms are checked multiple times.

Throughout this work, we have emphasized the importance of sharing in order to avoid repeated evaluations.  However, we only considered the sharing of sub-terms via lazy evaluation.  Other graph reduction techniques support the sharing of more than just lambda-terms, such as Lamping's optimal reduction algorithm \cite{DBLP:conf/popl/Lamping90} \cite{DBLP:conf/popl/GonthierAL92} \cite{DBLP:journals/jfp/AspertiGN96} and the atomic lambda-calculus \cite{DBLP:conf/fossacs/SherrattHGP20}.  It would be interesting to study the usability of these advanced graph reduction techniques in the context of convertibility checking.

\section*{Data-availability statement}

The Rocq development described in \secref{s:rocq-proof} and the benchmarks described in \secref{s:experimental} are available 
at \url{https://doi.org/10.5281/zenodo.17347533} (for reproduction)
and at \url{https://github.com/Ekdohibs/efficient-convertibility/} (for reuse).

\nocite{courant_2025_17347533}
\bibliographystyle{ACM-Reference-Format}
\bibliography{biblio}

\end{document}